# Quantized conductance in a CVD-grown nanoribbon with hidden Rashba effect

Jianfei Xiao,[1,2,*] Yiwen Ma,[1,2,*] Congwei Tan,[3,*,†] Kui Zhao,[1,4] Yunteng Shi,[1,2] Bingbing Tong,[1,5] Peiling Li,[1,5] Ziwei Dou,[1] Xiaohui Song,[1,5] Guangtong Liu,[1,5] Jie Shen,[1] Zhaozheng Lyu,[1,5] Li Lu,[1,2,5,‡] Hailin Peng,[3,§] and Fanming Qu[1,2,5,‖]

[1]Beijing National Laboratory for Condensed Matter Physics, Institute of Physics, Chinese Academy of Sciences, Beijing 100190, China

[2]University of Chinese Academy of Sciences, Beijing 100049, China

[3]Center for Nanochemistry, Beijing National Laboratory for Molecular Sciences, College of Chemistry and Molecular Engineering, Peking University, Beijing 100871, China.

[4]Beijing Key Laboratory of Fault-Tolerant Quantum Computing, Beijing Academy of Quantum Information Sciences, Beijing 100193, China

[5]Hefei National Laboratory, Hefei 230088, China

[*]These authors contributed equally to this work.

[†]Contact author: tancw-cnc@pku.edu.cn

[‡]Contact author: lilu@iphy.ac.cn

[§]Contact author: hlpeng@pku.edu.cn

[‖]Contact author: fanmingqu@iphy.ac.cn

**ABSTRACT**. Quantized conductance in quasi-one-dimensional systems not only provides a hallmark of ballistic transport, but also serves as a gateway for exploring quantum phenomena. Recently, a unique hidden Rashba effect attracts tremendous attention, which arises from the compensation of opposite spin polarizations of a Rashba bilayer in inversion symmetric crystals with dipole fields, such as bismuth oxyselenide ($Bi_2O_2Se$). However, investigating this effect utilizing conductance quantization is still challenging. Here we report the conductance quantization observed in a chemical vapor deposition (CVD)-grown high-mobility $Bi_2O_2Se$ nanoribbon, where quantized conductance plateaus up to $44 \cdot 2e^2/h$ ($e$ is the elementary charge, $h$ is the Planck's constant, and the factor 2 results from spin degeneracy) are achieved at zero magnetic field. Due to the hidden Rashba effect, the quantized conductance remains in multiples of $2e^2/h$ without Zeeman splitting even under magnetic field up to 12 T. Moreover, within a specific range of magnetic field, the plateau sequence exhibits the Pascal triangle series, namely $(1,3,6,10,15\ ...) \cdot 2e^2/h$, reflecting the interplay of size quantization in two transverse directions. These observations are well captured by an effective hidden Rashba bilayer model. Our results demonstrate $Bi_2O_2Se$ as a compelling platform for spintronics and the investigation of emergent phenomena.

## INTRODUCTION

The observation of quantized conductance, in units of $2e^2/h$ where $e$ is the elementary charge, $h$ is the Planck's constant, and the factor 2 originates from the spin degeneracy, serves as a hallmark of ballistic transport in quasi-one-dimensional (quasi-1D) systems. Moreover, conductance quantization also plays an important role in investigation of quantum phenomena, such as correlation effects [1], and Kronig-Penney model [2]. Quantized conductance was first achieved in quantum point contacts (QPCs) defined on III-V two-dimensional electron gases (2DEGs) [3–6], leveraging the high mobility and gate tunability, followed by metallic break junctions [7,8] and other two-dimensional (2D) systems [9–13]. Subsequently, benefiting from the improvements in material quality, quantized conductance was observed in ultraclean individual 1D systems [14–17], e.g., nanowires, nanoribbons, and nanotubes, but reported quantized plateaus are mostly limited under an index of 10 (namely, quantized conductance plateau $G \leq 10 \cdot 2e^2/h$).

Recently, a novel 2D semiconductor $Bi_2O_2Se$ has garnered significant attention [18–22] due to its outstanding electronic properties, including a moderate bandgap, small effective mass, ultrahigh carrier mobility, and strong spin-orbit coupling (SOC), together with the responsiveness to diverse external stimuli. Crucially, its excellent air stability and native oxide high-κ dielectric [23,24] $Bi_2SeO_5$ further enhance its potential for high-performance 2D electronic devices, including 2D-FinFET [18] and gate-all-around FET (GAAFET) [25]. Notably, the former investigations were mainly based on 2D $Bi_2O_2Se$. Herein, we choose quasi-1D $Bi_2O_2Se$ nanoribbon as the platform to investigate ballistic transport approaching high quantized indices, particularly involving hidden Rashba effect and Pascal triangle series, as detailed in the following.

First, due to the large Fermi velocity [26], suppressed electron back-scattering [22], and the self-modulation doping [27], $Bi_2O_2Se$ gains a remarkably high mobility [21]. Interestingly, the self-modulation doping originates from that the conductive $[Bi_2O_2]_n^{2n+}$ layers are spatially separated from the defect-rich $[Se]_n^{2n-}$ layers, minimizing impurity scattering. This exceptional high-mobility has facilitated the demonstration of quantum Hall effect (QHE) [28,29]. Second, because of the inversion symmetry of $Bi_2O_2Se$ and the presence of an intrinsic dipole field, hidden Rashba effect [30] emerges. This effect provides efficient electrical tunability for spin FET [31,32], and has been probed by angular resolved photoemission spectroscopy and photocurrent in, e.g., $WSe_2$ [33,34], BISCCO(2212) [35], ZrSiTe [36], and GdSbTe [37], while evidence from electrical transport experiment is still lacking. Furthermore, the hidden Rashba effect with spin-layer locking exhibits a distinct spin texture that maintains spin degeneracy even under high magnetic fields [28], facilitating quasi-1D ballistic transport as a probe for this phenomenon. Third, size quantization in a ballistic system along two transverse directions under the harmonic-oscillator-type parabolic potentials induces a complex evolution of conductance quantization, which could follow a characteristic pattern of the Pascal triangle series [1,38]. Considering the electron-electron interaction [39], $Bi_2O_2Se$ serves as a promising platform for investigating correlation effects [1].

In this Letter, we report the conductance quantization behavior, combined with the hidden Rashba effect, in high-quality chemical vapor deposition (CVD)-grown $Bi_2O_2Se$ nanoribbons. Quantized conductance plateaus up to $44 \cdot 2e^2/h$ are observed, the highest in individual 1D systems. In particular, owing to the hidden Rashba effect resulted from the unique Rashba bilayer structure, the conductance quantization remains at multiples of $2e^2/h$—rather than Zeeman-splitting-induced $e^2/h$—even under magnetic fields up to 12 T. Furthermore, within a range of magnetic field, the quantized conductance plateaus exhibit a sequence corresponding to the Pascal series. These experimental results are well reproduced by theoretical calculations based on an effective hidden Rashba bilayer model.

## RESULTS

$Bi_2O_2Se$ nanoribbons are grown by CVD method [40,41], and the two-terminal devices are fabricated by standard electron-beam lithography techniques. Figure 1(a) depicts a schematic of the device, where a $Bi_2O_2Se$ nanoribbon is contacted by the source and drain, with a channel length of $L$≈550 nm. Notably, $Bi_2O_2Se$ features an inversion-symmetric crystal structure, where tetragonal $[Bi_2O_2]_n^{2n+}$ and $[Se]_n^{2n-}$ layers are staggered and stacked along the c-axis [42]. As schematically depicted in the left inset of Fig. 1(a), the two Bi monolayers in the $[Bi_2O_2]_n^{2n+}$ layers form a peculiar hidden Rashba bilayer due to opposite interlayer polarizations [28] (labelled by $P_{\mathrm{in}}$ in different colors). The alternative dipole fields break the local inversion symmetry of the monolayers, and give rise to opposite strong Rashba effect with compensated spin textures, which results in the hidden Rashba effect [30]. In this scenario, as illustrated in the right inset of Fig. 1(a), each band is two-fold degenerate, with an expected extremely small effective $g$-factor, indicating the suppression of Zeeman splitting under high magnetic fields.

Figure 1(b) depicts the confining potential of spin-degenerate electrons, which come from different monolayers of the hidden Rashba bilayer. Notably, the perpendicular magnetic field along the $z$ direction modulates the electrostatic confining potentials in the $y$ direction [43], and thus the size quantization in the nanoribbon, effectively. The lowest several sublevels denoted by $\mathrm{n_i} = 0, 1, 2, \ldots$ (where $\mathrm{i} = \mathrm{y}, \mathrm{z}$) are the quantized states in $y$ and $z$ directions, respectively. Figure 1(c) presents the atomic force microscopy (AFM) image of the device, with a thickness of 50 nm. As shown in Fig. 1(d), as a result of the size quantization in the two transverse directions of the nanoribbon, the quantized conductance under a range of magnetic fields follows the characteristic sequence of $G = \frac{N(N+1)}{2} \cdot 2e^2/h$ (where $N = 1,2,3, \ldots$), corresponding to the purple diagonal in Pascal triangle. The hidden Rashba effect results in the spin degenerate bands, leading to the persistence of quantized conductance being multiples of $2e^2/h$.

To demonstrate the quantized conductance behavior of our $Bi_2O_2Se$ nanoribbon device, we performed low-temperature ($T$ = 1.5 K) transport measurements under perpendicular magnetic fields along the $z$-direction (Fig. 1(b)). Note that a series resistance of 350 Ω, determined by the deviation of the first plateau at $B$ = 12 T from the expected conductance of $2e^2/h$, is subtracted from all measured resistance data. In addition, all the conductance traces were measured at zero d.c. bias voltage, unless otherwise stated.

At $B$ = 0 T (Fig. 2(a)), a series of quantized conductance plateaus with indices reaching 44 (in units of $2e^2/h$) are observed as the gate voltage $V_{\mathrm{G}}$ varies. Noteworthy, the index of 44 represents the highest value among individual 1D systems, where a maximum index of 10 was reported, as illustrated in Fig. 2(e) (see Supplemental Material [44]). Even taking QPCs defined on 2D systems into consideration [3,9,45], our observation of 44 is still among the best. From the fitting of the $V_{\mathrm{G}}$-dependent conductance ($G$) curve in Fig. 2a, we extracted the field-effect mobility $\mu \sim 1.11 \times 10^4\ \mathrm{cm^2V^{-1}s^{-1}}$ (see Supplemental Material [44]). In addition, the absence of some certain plateaus in Fig. 2(a) can be generally attributed to the degenerate subbands as shown by the schematic of the Fermi level alignment conditions in Fig. 2(b). In other words, the transitions between conductance plateaus occur when the Fermi level aligns with the bottoms of subbands, whose degeneracy is governed by the confining potentials. It is crucial to point out that each parabolic band contributes a quantized conductance of $e^2/h$. When $B$ is increased to 9.75 T, the conductance trace as a function of $V_{\mathrm{G}}$ (Fig. 2(c)) exhibits two remarkable characteristics. First, the quantized conductance remains in units of $2e^2/h$, indicating the suppression of Zeeman splitting because of the hidden Rashba effect. Second, the unconventional quantized plateau series that follows the Pascal triangle, i.e., $1{,}3{,}6{,}10{,}15 \cdot (2e^2/h)$, is observed, implying a unique band degeneracy feature as schematically depicted in Fig. 2(d), which will be discussed later.

We then apply a d.c. bias voltage $V_{\mathrm{b}}$ to investigate the spectroscopy of the conductance quantization. Figure 3 presents the spectroscopy measurement at $B$ = 12 T, revealing a series of quantized conductance of $1{,}3{,}6 \cdot (2e^2/h)$, which is also in well conformity with the highlighted diagonal of the Pascal triangle in Fig. 1(d). We plot a series of conductance $G$ as a function of $V_{\mathrm{b}}$ with different $V_{\mathrm{G}}$ in Fig. 3(a). To achieve a clearer understanding, we present the normalized transconductance $\mathrm{d}G/\mathrm{d}V_{\mathrm{G}}$ as a function of $V_{\mathrm{G}}$ and $V_{\mathrm{b}}$ in Fig. 3(b), where the transitions between the quantized conductance plateaus are represented by the bright regions. Furthermore, the diamond-shaped regions outlined by dashed lines connecting the circles correspond to the conductance plateaus quantized in multiples of $2e^2/h$. In addition, these quantized plateaus, labelled with A, B, and C, align with the clustering of the conductance curves in Fig. 3(a).

In Fig. 3(c) the Fermi level alignments noted by A, B, and D, where $\mu_s$ and $\mu_D$ denote the chemical potentials of the source and drain, respectively, illustrate the corresponding conditions in Figs. 3(a), (b). In addition, conductance traces measured at different magnetic fields (see Supplemental Material [44]) consistently persist the conductance quantization in units of $2e^2/h$. Moreover, the transconductance spectroscopy under different magnetic fields (see Supplemental Material [44]) reveal that some certain quantized plateaus (e.g., $2G_0, 3G_0, 4G_0, ...$) gradually emerge and/or disappear as $B$ increases, indicating an unusual magnetic field dependence of the subband evolution.

To explore the evolution of quantized conductance with the magnetic field ($B$), we plot the conductance traces as a function of $V_G$ in Fig. 4(a), under magnetic fields ranging from 0 to 12 T in steps of 0.25 T. The indices of the quantized conductance plateaus are labelled (demonstration of individual conductance curves, along with alternative versions of the conductance map are provided in Supplemental Material [44]). Notably, under all the magnetic fields up to 12 T, the half-integer plateaus (including 0.5 and $1.5 \cdot 2e^2/h$) are absent in all the conductance traces in Fig. 4(a), reflecting the suppression of Zeeman splitting due to the hidden Rashba effect, as can also be seen in Fig. 2(b) and in the spectroscopy of Fig. 3(b).

To investigate the evolution of the quantized conductance plateaus, in Figs. 4(b) and 4(c), we present the renormalized transconductance ($\mathrm{d}G/\mathrm{d}\mu$) and conductance maps as a function of $B$ and chemical potential $\mu$, respectively (the transformation from $V_G$ to $\mu$, i.e., the gate lever arm, is explained in Supplemental Material [44]). Particularly, in Fig. 4(b), the quantized conductance plateaus are represented by the dark purple regions, some of which are labelled with indices, whereas the bright boundaries delineating these regions stand for the evolution of subbands induced by size quantization along the two transverse directions. The evolution of the subbands are traced in Fig. 4(b), and then they can be categorized and grouped based on their slopes with respect to $\mu$ under high magnetic fields, as shown by the solid lines with different colors in Fig. 4(c) and detailed in Supplemental Material [44]. By tracking the origin of each subband at low magnetic fields, we find that they arise from energy quantization in both $y$ and $z$ directions: subbands of different colors correspond to quantization due to the finite thickness along the $z$ direction, whereas those of the same color stem from the constrained width along the $y$ direction. Consequently, we label the subbands in the form of $|n_z, n_y\rangle$ in Fig. 4(c) and Fig. S11 [44], where $n_y$ and $n_z$ denote the quantum numbers associated with size quantization in $y$ and $z$ directions, respectively.

To capture the quantized conductance behavior, we establish an effective hidden Rashba bilayer model considering confining potentials in both $y$ direction, which is modulated by the magnetic field, and $z$ direction. In fact, under zero magnetic field, the level spacings between subbands quantized along the $y$ direction—indicated by same colors in Fig. 4(c) corresponding to the same $n_z$ but varying $n_y$—are nearly identical. Similarly, the subbands resulting from quantization in the $z$ direction, namely those represented by different colors with fixed $n_y$ and different $n_z$, exhibit identical spacings as well. This naturally suggests that the confining potentials in $y$ and $z$ directions can both be well approximated by a harmonic oscillator potential, as the quantized energy levels possess equidistant spacing described by $E_\xi = (\xi + 1/2)\hbar\omega$ (where $\omega$ is the eigenfrequency, $\xi$ is the energy level index). Thus, the effective Hamiltonian for the system can be expressed as: $H = \frac{(\boldsymbol{p}+e\boldsymbol{A})^2}{2m'} + \frac{1}{2}g'\mu_B B\sigma_z + \frac{1}{2}m'\omega_y^2 y^2 + \frac{1}{2}m'\omega_z^2 z^2$ , where $m'$ and $g'$ are the renormalized effective mass and $g$-factor from the ab initio calculation based on hidden Rashba bilayer model involving multi-bilayers, respectively, $\mu_B$ denotes the Bohr magneton, σ is the Pauli matrix, $\boldsymbol{A} = A_x\,\boldsymbol{e}_x = -By\,\boldsymbol{e}_x$ represents the vector potential, $\omega_{y,z}$ denotes the eigenfrequency of the confining potentials in $y$ and $z$ directions, respectively.

Such effective model describes the subbands with the following form:

$$E(n_y, n_z; B) = \hbar\Omega\left(n_y + \frac{1}{2}\right) + \hbar\omega_z\left(n_z + \frac{1}{2}\right) \pm \frac{1}{2}g'\mu_B B, \qquad (1)$$

where $\Omega = \sqrt{\omega_c^2 + \omega_y^2}$, $\omega_c = eB/m'$ is the cyclotron frequency. Due to the unavoidable effects of finite temperature and measurement noise, energy broadening occurs, resulting in the smearing of the density of states, which may hinder the differentiation of Zeeman splitting under magnetic fields up to 12 T. Here, we assume a Gaussian broadening $W(\mu; \varepsilon_0, \Gamma) = \frac{1}{\sqrt{2\pi}\Gamma}\,e^{-(\mu-\varepsilon_0)^2/(2\Gamma^2)}$ with $\Gamma$ = 0.7 meV determined from the Gaussian fitting of the measured spectroscopy (see Supplemental Material [44]).

To obtain $m'$ and $g'$ on the basis of the hidden Rashba effect together with interlayer coupling, we employ an ab initio calculation for a 40-unit-cell-thick $Bi_2O_2Se$ nanoribbon based on the thickness of 50 nm (detailed in Supplemental Material [44]). Herein, the Rashba SOC strength $\alpha$ alternates in sign between every two $[Bi_2O_2]_n^{2n+}$ layers along the $z$-axis, mimicking a Su-Schrieffer-Heeger chain [46]. The orbital effects induced by the magnetic field are implemented via Peierls substitution, $\boldsymbol{p} \to \boldsymbol{p} + e\boldsymbol{A}$. Subsequently, through an exact unitary transformation, the Hamiltonian can be expressed under the basis of Landau levels [47] as:

$$H_{nn'}^{ll'} = \hbar\frac{eB}{m^*}\left(n + \frac{1}{2}\right)\delta_{nn'}\delta_{ll'}\sigma_0 + \frac{1}{2}g_e\mu_B B\delta_{nn'}\delta_{ll'}\sigma_z - t\delta_{nn'}\delta_{l,l'\pm1}\sigma_0$$
$$+(-1)^l\frac{-i\alpha}{\sqrt{2}\ell_B}\,\delta_{ll'}\left(\sqrt{n+1}\,\delta_{n,n'-1}\sigma_+ - \sqrt{n}\,\delta_{n,n'+1}\sigma_-\right), \qquad (2)$$

where $n$ is the quantum number of Landau levels, $l$ is the layer index, $\sigma_\pm = \sigma_x \pm i\sigma_y$, $\ell_B = \sqrt{\hbar/eB}$ is the magnetic length, $g_e = 2$ is the bare Landé $g$-factor, $t$ represents the interlayer coupling.

By taking $\alpha = 1.45\ \mathrm{eV \cdot Å}$ [28], we find that when $m^* = 0.08\, m_e$ and correspondingly $t = 0.32$ eV, the simulation agrees with the experimental data well, which gives renormalized $m' = 0.084\, m_e$ and $g' = 0.78$. Notably, the suppression of $g'$ to 0.78 is induced by the hidden Rashba effect and interlayer hybridization (see Supplemental Material [44]). Specifically, the spin-conserved interlayer coupling of electrons hybridizes the subbands with opposite spin textures originating from monolayers with $+\alpha$ and $-\alpha$, namely the hidden Rashba effect. This effect further competes with the bare Zeeman effect, which leads to the suppressed $g'$. Therefore, Zeeman splitting becomes undetectable even under 12 T. As shown in Fig. 4(d), the simulation exhibits good agreement with the experimental observations, enabling the conductance quantization as a reliable method for investigating this effect. In addition, the shift of high-degeneracy points (i.e., subband crossings) towards lower magnetic fields in Fig.4(b), compared to the simulation, can be attributed to the derivation of the confining potential from the ideal harmonic oscillator model.

Finally, to better understand the Pascal-like quantized conductance plateaus observed in high magnetic fields, we begin with Eq. (1), which demonstrates that the subband energy can be simplified as a linear combination $\eta n_y + \zeta n_z$, where $\eta$ and $\zeta$ represent the coefficients. The fact that the magnetic field modulates the confining potential in $y$ direction, effectively results in a corresponding adjustment of $\eta$. The index of the conductance plateau at a specific chemical potential $\mu_0$ is actually the number of subbands intersecting the Fermi level, as determined by the count of natural number solutions to the inequality $\eta n_y + \zeta n_z \leq \mu_0$. For instance, in Fig. 4(b), $\eta = \zeta$ is satisfied around 10 T, and the amount of natural number solutions (subbands below the Fermi level) to $n_y + n_z \leq \mathrm{Const.}$ defines the conductance plateau indices, which follow a series corresponding to Pascal triangle with the transverse dimension of 2. As a result, $G = \frac{\mathrm{N(N+1)}}{2} \cdot 2e^2/h$, namely $G = 1{,}3{,}6{,}10{,}15 \cdot (2e^2/h)$, as observed experimentally. Additionally, at about 4 T, distinct diamond-like patterns emerge in Fig. 4(b), consistent with the solution of $n_y/2 + n_z \leq \mathrm{Const.}$ Notably, due to the confinement of harmonic-oscillator-type potentials in both the $y$ and $z$ directions subbands evolution in Fig. 4(b) exhibits a unique moiré pattern.

## CONCLUSIONS

In conclusion, we demonstrate the experimental observation of ballistic transport in a gate-tunable device based on CVD-grown $Bi_2O_2Se$ nanoribbon. Notably, the highest quantized conductance plateau index reaches 44 at zero magnetic field, which was previously unachievable in this context. Particularly, the quantized conductance maintains to be multiples of $2e^2/h$ even under magnetic fields up to 12 T, because of the unique hidden Rashba effect originating from the distinct Rashba bilayer structure. This also demonstrates conductance quantization as an efficient method for probing such hidden Rashba effect. Furthermore, due to the size quantization in both $y$ and $z$ directions together with the modulation by the magnetic field, the observed quantized conductance sequence under a specific range of magnetic fields follows the Pascal triangle series. Finally, by means of an effective hidden Rashba bilayer model we demonstrate spectra that agree well with our experiment.

The observation of ultrahigh quantized conductance plateaus serves as a clear signature of ballistic transport, where disorder represents a significant barrier to the practical implementation of quasi-1D materials in applications such as Majorana zero modes [48]. Furthermore, the spin-layer locking, manifested by the unique hidden Rashba effect, positions $Bi_2O_2Se$ as a promising candidate for investigating unique properties [30], which were thought to only exist in inversion-asymmetric systems, including piezoelectricity and second-harmonic generation. Moreover, its exceptional electrical tunability and high mobility reinforce its potential for spintronic applications, such as electrically controlled spin FET [31,32]. Beyond its relevance to applications, $Bi_2O_2Se$ also provides a fertile ground for exploring emergent phenomena, including the spin Hall effect and nonlinear physics, and could be a compelling platform to investigate correlation effects [1] considering the electron-electron interaction [39].

## ACKNOWLEDGMENTS

This work was supported by the National Key Research and Development Program of China (2022YFA1403400); by the National Natural Science Foundation of China (92065203, 92365207, and 22432001); by the Strategic Priority Research Program of Chinese Academy of Sciences (XDB33000000); by the Synergetic Extreme Condition User Facility sponsored by the National Development and Reform Commission; and by the Innovation Program for Quantum Science and Technology (2021ZD0302600).

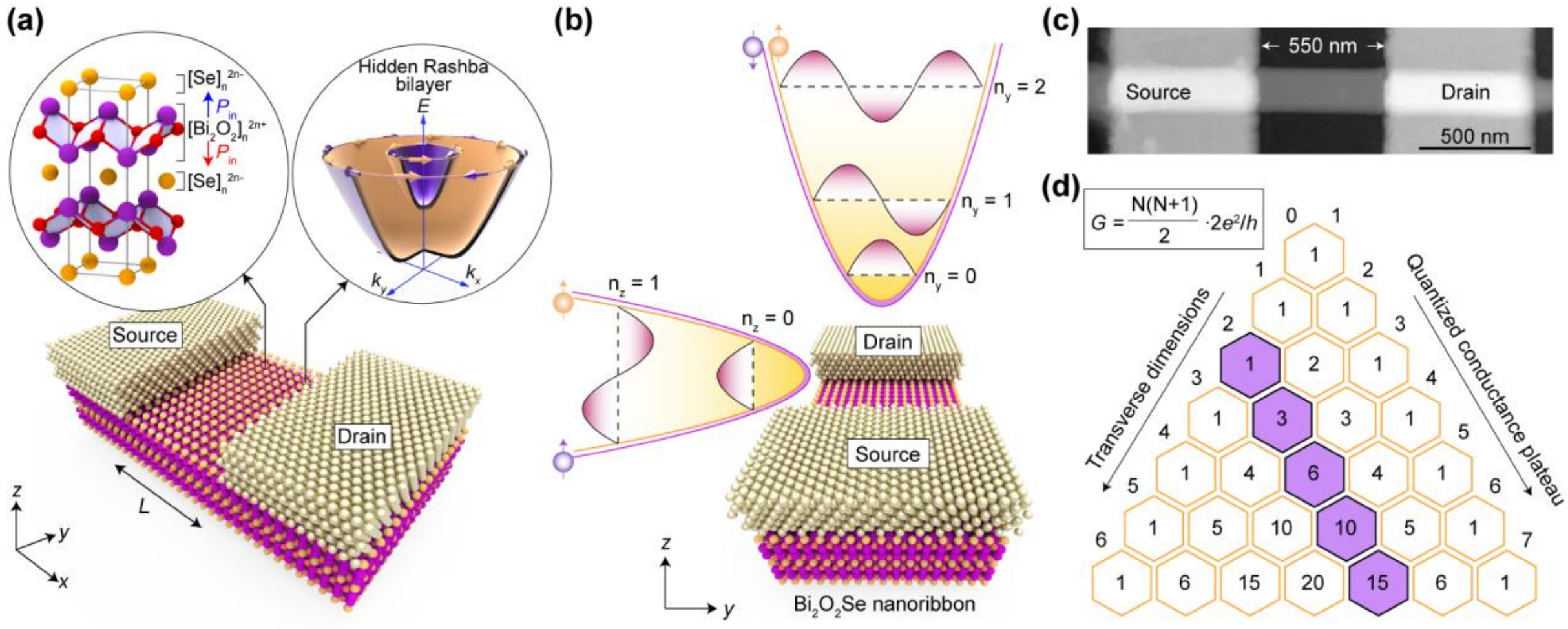

FIG. 1. (a) Schematic of the $Bi_2O_2Se$ nanoribbon device. The left inset illustrates the layered crystal structure. The right inset depicts the hidden Rashba effect arising from the unique Rashba bilayer structure. The bands in different colors represent different spin textures, while the arrows stand for the spin orientations. (b) Illustration of the confining potential, identical for the two spin textures. For clarity, the two confining potentials for two degenerate spins are slightly separated. (c) AFM image of the device. (d) Pascal triangle of the quantized conductance, described by $G = \frac{N(N+1)}{2} \cdot 2e^2/h$ for the highlighted diagonal with the transverse dimension of 2.

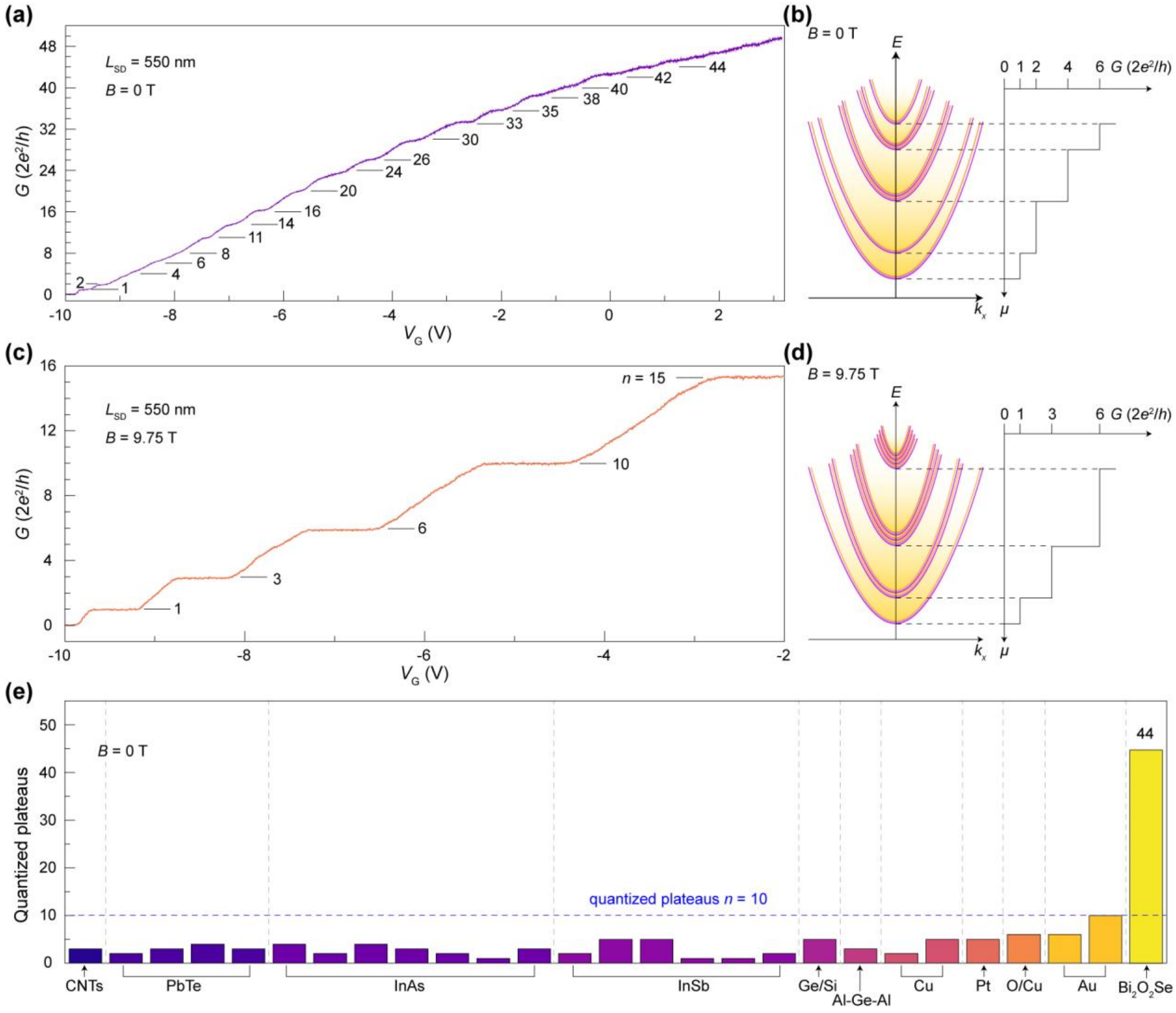

FIG. 2. (a) Quantized conductance ($G$), expressed in units of $2e^2/h$ as a function of gate voltage $V_G$ at $B$ = 0 T and $T$ = 1.5 K. (c) Quantized conductance at $B$ = 9.75 T, which exhibits a novel degeneracy of quantized conductance steps following the Pascal triangle sequence. (b), (d), Fermi level alignments corresponding to the quantized conductance observed in (a) and (c), respectively. (e) Survey of quantized plateau index among individual 1D materials.

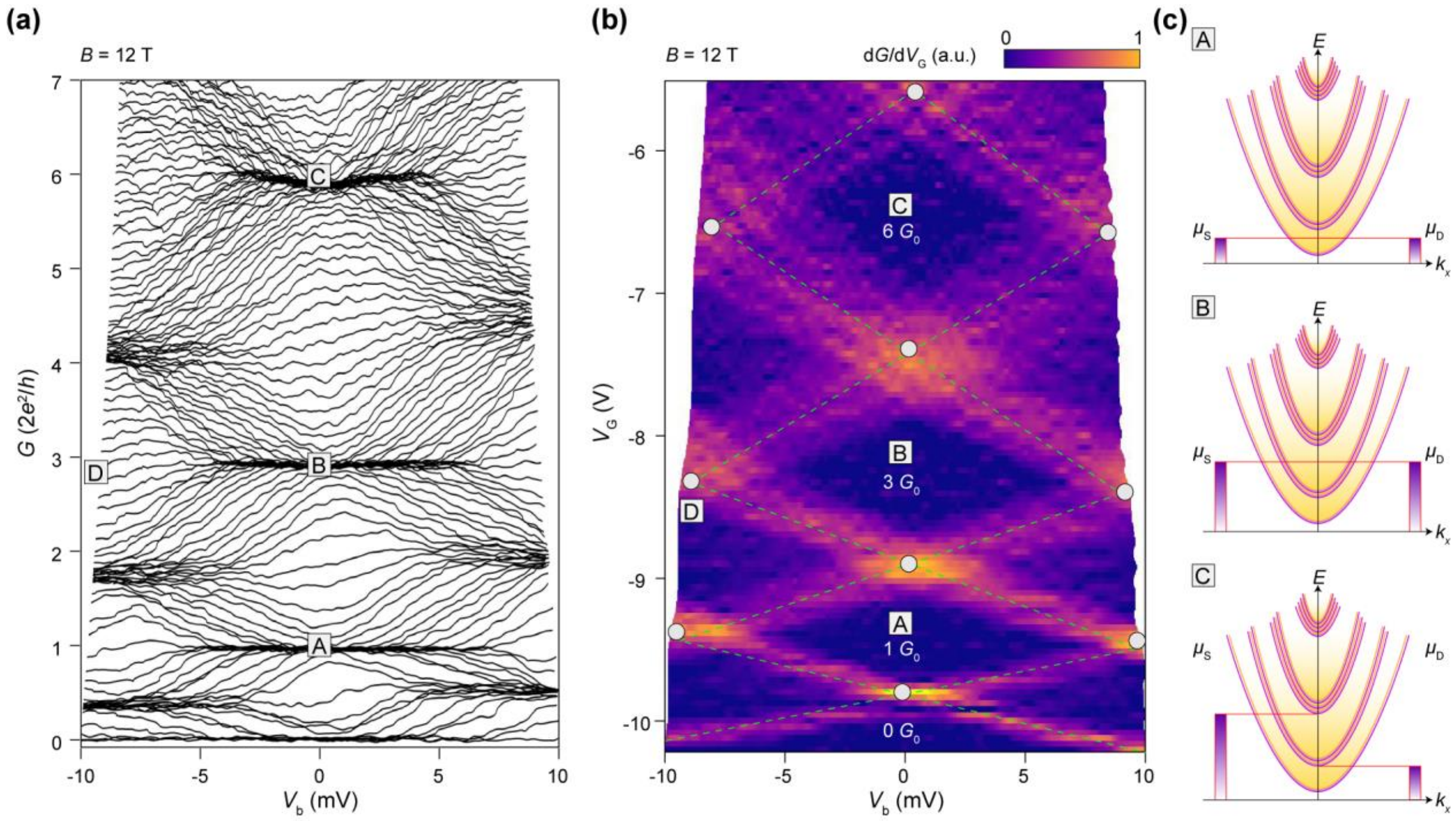

FIG. 3. (a) Conductance $G$ as a function of d.c. bias voltage $V_\mathrm{b}$ at 12 T for gate voltage $V_\mathrm{G}$ between -10.1 V and -5.5 V (from bottom to up). (b) Normalized transconductance ($\mathrm{d}G/\mathrm{d}V_\mathrm{G}$) map as a function of $V_\mathrm{G}$ and $V_\mathrm{b}$, where $G_0 = 2e^2/h$. (c) Fermi level alignment conditions, labelled by A, B, and D in (a) and (b), respectively.

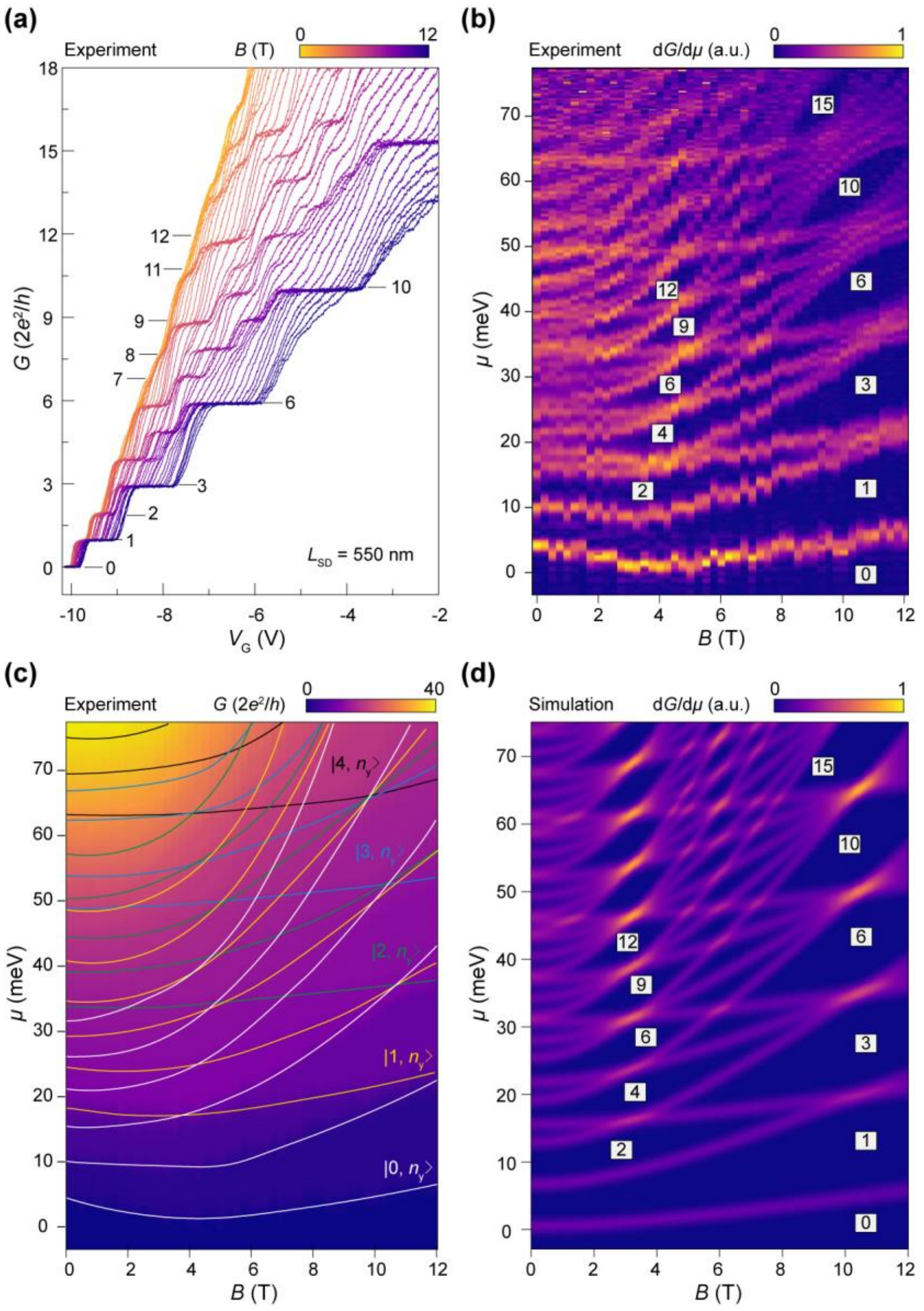

FIG. 4. (a) Quantized conductance evolution under different magnetic fields plotted as a function of $V_{\mathrm{G}}$. (b), (c) Renormalized transconductance ($\mathrm{d}G/\mathrm{d}\mu$) and conductance ($G$) maps plotted against $\mu$ and $B$, respectively. (d) Simulation of $\mathrm{d}G/\mathrm{d}\mu$ based on the effective model (Eq. (2)), demonstrating consistency with (b) and (c), considering a broadening of 0.7 meV. The numbers label the quantized conductance indices, while $|n_z, n_y\rangle$ denotes the subbands.

**Supplemental Material for**

# Quantized conductance in a CVD-grown nanoribbon with hidden Rashba effect

Jianfei Xiao,[1,2,*] Yiwen Ma,[1,2,*] Congwei Tan,[3,*,†] Kui Zhao,[1,4] Yunteng Shi,[1,2] Bingbing Tong,[1,5] Peiling Li,[1,5] Ziwei Dou,[1] Xiaohui Song,[1,5] Guangtong Liu,[1,5] Jie Shen,[1] Zhaozheng Lyu,[1,5] Li Lu,[1,2,5,‡] Hailin Peng,[3,§] and Fanming Qu[1,2,5,‖]

## A. Methods

### *1. Sample growth and characterization*

The CVD-grown $Bi_2O_2Se$ nanoribbons are obtained by transferring the as-prepared vertical $Bi_2O_2Se$ fins directly onto $SiO_2$/Si substrates. Vertical $Bi_2O_2Se$ fins were synthesized in homemade CVD systems equipped with 110-cm horizontal tube furnaces and quartz tubes with 1-inch internal diameters. $Bi_2O_3$ powder (Alfa Aesar, 99.999%) and $Bi_2Se_3$ powder (Alfa Aesar, 99.999%) were utilized as co-evaporation precursors. High-purity Ar gas served as carrier gases. Following the vapor-phase growth, the synthesis procedure for vertical $Bi_2O_2Se$ fins, as an extension of the previously reported methodology [22,40], is as follows. $Bi_2O_3$ powder was positioned at the hot center of the furnace, while $Bi_2Se_3$ powder was located 4 cm upstream with source temperature 615-645 $^oC$. High-purity Ar gas was supplied at a typical flow rate of 100 sccm (where sccm denotes standard cubic centimeters per minute), and the total system pressure was maintained at 400 Torr. Single-crystal $LaAlO_3$ (100) substrates were placed on top of the intersection of the $Bi_2O_3$ and $Bi_2Se_3$ precursors with an approximate gap of 5 mm. The growth time was 1 min.

### *2. Device fabrication*

The device utilized in this study was fabricated using standard electron-beam lithography (EBL) techniques based on the CVD-grown high-quality $Bi_2O_2Se$ nanoribbon. Initially, a Ti/Au (with a thickness of 5/15 nm) bottom gate with a width of 8 μm and a length of 20 μm was patterned on a 300 nm $SiO_2$/Si substrate by EBL (Raith eLine Plus) and electron-beam deposition (AD Nano). Then a high-κ dielectric hBN nanoflake, with a thickness of 18 nm obtained via mechanical exfoliation with the assistance of PDMS substrate, was transferred onto the bottom gate utilizing a home-built transfer station under the optical microscope, aided by propylene carbonate (PPC) gel droplets. After immersing in Acetone for 6 hours, thermal annealing at 150 ℃ in an ozone atmosphere was carried out to remove residual PPC, and the $Bi_2O_2Se$ nanoribbon with a thickness of 50 nm and a width of 140 nm was precisely positioned onto the hBN using the same transfer station. To ensure complete removal of PPC, the sample underwent a second acetone soak for 10 hours. Finally, after a second EBL step, Ti/Au electrodes (5/90 nm) were deposited, contacting both the nanoribbon and the bottom gate. All evaporation processes were performed after in-situ Argon plasma etching to ensure a clean interface. The AFM image of the device was performed on a Bruker Nanoscope system (Dimension Edge).

### *3. Electrical transport measurements*

All low-temperature electrical transport measurements were carried out in a cryofree static superconducting magnet system (TeslatronPT, Oxford Instruments) in a quasi-four-terminal configuration. Standard lock-in techniques were employed to measure the differential conductance at a low-noise level. Typically, the current ($I_{\mathrm{a.c.}}$) was collected by lock-in amplifiers (NF LI 5650) with an a.c. excitation voltage ($V_{\mathrm{a.c.}}$) of 75 μV at a low frequency of 30.9 Hz. The gate and d.c. bias voltages were applied by a two-channel source meter (Keithley 2612B). Notably, the conductance is given by $G = I_{\mathrm{a.c.}}/V_{\mathrm{a.c.}}$.

## B. Survey of quantized plateau index

This section presents a survey of the quantized plateau index observed at magnetic field $B$ = 0 T. The reported data in literature, as shown in Fig. 2(e), has been reorganized into Table SI. Here, we focus on individual 1D materials, e.g., nanowires, nanoribbons, and nanotubes, excluding the nano-constrictions defined on 2DEGs. The materials are categorized into two classes: semiconductors, which include carbon nanotube (CNT) [49], PbTe [50–53], InAs [54–60], InSb [16,61–65], Ge/Si [17], and Al-Ge-Al [66], and metals, such as Cu [67,68], Pt [68], Au [69,70], and oxygen covered Cu [68].

## C. Extraction of mobility

This section details the extraction of the field-effect mobility of the $Bi_2O_2Se$ nanoribbon. Fig. S1(a) displays the full-range gate-dependent conductance curve of the device at magnetic field $B$ = 0 T. The observed negative pinch-off voltage ($V_{\mathrm{th}}$) indicates inherent n-type doping in the semiconducting nanoribbon. Field-effect mobility ($\mu$) was determined through quantitative analysis of the transfer characteristics using a modified mobility model that accounts for contact resistance effects [71],

$$G(V_{\mathrm{G}}) = \left(R_{\mathrm{c}} + \frac{L^2}{\mu C_{\mathrm{G}}(V_{\mathrm{G}} - V_{\mathrm{th}})}\right)^{-1}, \tag{S1}$$

where $C_{\mathrm{G}}$ is the capacitance, $R_{\mathrm{c}}$ denotes the contact resistance given by the fitting, and $L$ is the channel length.

To ensure accuracy, the gate capacitance $C_{\mathrm{G}}$ was calculated via two-dimensional finite element method (2D-FEM) simulations using COMSOL Multiphysics®. For a channel length of 550 nm, the simulation yielded $C_{\mathrm{G}}$ ~ 0.13 fF. Fitting of the experimental conductance curve using Eq. (S1) (orange dashed line in Fig. S1(a)) gives an extracted mobility of $\mu \sim 1.1 \times 10^4\ \mathrm{cm^2V^{-1}s^{-1}}$. Furthermore, the exceptionally high mobility in $Bi_2O_2Se$ can be attributed to its small effective mass, large Fermi velocity, and the unique spatially segregated conduction mechanism [21]. Specifically, the conductive $[Bi_2O_2]_{\mathrm{n}}^{2\mathrm{n}+}$ layers are electronically decoupled from adjacent Se-vacancy rich $[\mathrm{Se}]_{\mathrm{n}}^{2\mathrm{n}-}$ layers. This structural configuration effectively suppresses impurity scattering, thereby enhancing carrier mobility. In addition, $G$ with $R_{\mathrm{s}}$ subtracted was demonstrated as a function of $V_{\mathrm{G}}$ under zero magnetic field in a full-scale in Fig. S1(b), as also partially illustrated in Fig. 2(a). Notably, quantized conductance plateau indices that are higher than 44, which we did not emphasize in the main text, could be recognized as well.

## D. Magnetic field evolution of quantized conductance

This section presents supplementary data on conductance traces under different magnetic fields (Fig. S2), in addition to the traces at $B$ = 0, 9.75 T shown in the main text. Notably, the series resistance of $R_{\mathrm{s}}$ = 350 Ω was determined by the deviation of the first plateau from $2e^2/h$ at 12 T. All reported conductance traces presented in the main text and supplementary material (except for Fig. S1(a)) were renormalized by subtracting $R_{\mathrm{s}}$.

Key characteristics of the conductance traces depicted in Fig. S2 are as follows. 1) Enhanced plateau quantization is observed when the magnetic field increases, attributable to the suppressed back-scattering and magnetic depopulation through the orbital effect. 2) The quantized plateaus remain to be multiples of $2e^2/h$ under magnetic field up to 12 T, and half-integer quantized conductance plateaus, such as 0.5 and $1.5 \cdot 2e^2/h$, are absent, due to spin degeneracy (the suppression of Zeeman splitting) provided by the unique hidden Rashba effect. 3) Some of the plateaus vanish when magnetic field increases, indicating the peculiar field evolution of the subbands.

**E. Bias spectroscopy under different magnetic fields**

In this section, supplementary d.c.-bias spectra are presented. In order to investigate the spectroscopy of our ballistic nanoribbon, we apply d.c. bias voltages and obtain the spectra under different magnetic fields, as depicted in Fig. 3 and Fig. S3. It is worth mentioning again that all the quantized conductance represented by the clusters of conductance curves in the series of conductance traces shown in Fig. 3(a) and S3(a) to F, remain in units of $2e^2/h$ even at $B$ = 12 T. This highlights the role of the hidden Rashba effect in preserving the spin degeneracy rather than Zeeman splitting. The transconductance spectra shown in Fig. S3(g-l) are in correspondence to the conductance traces shown in Fig. S3(a-f), respectively. For clarity, the numbers inside the diamond shaped regions in transconductance spectra (Fig. S3(g-l)) indicate the sequence of quantized conductance plateaus. In addition, the white dashed lines in Fig. S3(g), (h) highlight the quantized conductance diamonds for better visualization. It is noteworthy that, the quantized conductance sequence at $B$ = 10 T establishes the series of Pascal triangle, namely $(0,1,3,6,\ldots) \cdot 2e^2/h$.

**F. Further information for the evolution of quantized plateaus**

In this section, we present the extended data for quantized conductance in magnetic fields. In the following supplementary figures (Fig. S4 to S8), we demonstrate the subband evolution in magnetic field using different types of plotting according to the same date set of Fig. 4(a).

**G. Quantification of the gate lever arm**

In this section, we detail the extraction of the lever arm used to convert the gate voltage $V_{\mathrm{G}}$ into chemical potential $\mu$. Fig. S9(a) demonstrates the full-scale d.c.-bias spectroscopy at $B$ = 8 T, offering a comprehensive view of the transconductance evolution as a function of $V_{\mathrm{b}}$ and $V_{\mathrm{G}}$. The lever arm factor $f_{\mathrm{LA}}$ is determined from the slope of the subband features, expressed as $f_{\mathrm{LA}} = \partial V_{\mathrm{b}}/\partial V_{\mathrm{G}}$ at the transconductance peaks (at $V_{\mathrm{b}} = 0$ mV), as indicated by the dashed lines in Fig. S9(a). Notably, $f_{\mathrm{LA}}$ increases as the gate voltage decreases towards pinch-off. We extracted $f_{\mathrm{LA}}$ from the slope of the transconductance peaks near zero bias at different gate voltages, as depicted in Fig. S9(b).

To find a functional relationship between $V_{\mathrm{G}}$ and $\mu$, we assume the following equation:

$$ef_{LA} = \frac{\partial \mu}{\partial V_{\mathrm{G}}} = \lambda_1 \exp(-\lambda_2 V_{\mathrm{G}}) + \lambda_3, \tag{S2}$$

and it turns out to reach a good fitting of the data points in Fig. S9(b). Therefore, we obtain the relationship between $\mu$ and $V_{\mathrm{G}}$:

$$\mu = -\frac{\lambda_1}{\lambda_2} \exp(-\lambda_2 V_{\mathrm{G}}) + \lambda_3 V_{\mathrm{G}} + \mu_0, \tag{S3}$$

where the fitting parameters $\lambda_1 = 0.0134$ meV/V, $\lambda_2 = 0.7322\ \mathrm{V}^{-1}$, and $\lambda_3 = 6.1423$ meV/V. It is worth mentioning that Eq. S3 is employed for conversion of $V_{\mathrm{G}}$ into $\mu$, enabling the analysis of the subband evolution in the energy scale, such as in Fig. 4(b).

## H. Extended analysis of subband evolution

This section details the methodology for subband categorization. Based on the unlabeled transconductance map in Fig. S8, it is clear that the bright boundaries separating the quantized plateau regions (represented by dark purple), stand for the evolution of subbands. Notably, the evolution of subbands can be traced and identified. Subsequently, the subbands are depicted across the full magnetic field range (from 0 to 12 T), as illustrated by the dashed lines in Fig. S10, S11. By analyzing the slopes of the subbands with respect to chemical potential under high magnetic fields, we classify them into distinct groups and assign subband labels in the form $|n_z, n_y\rangle$, following the subband energy:

$$E(n_y, n_z; B) = \hbar\Omega\left(n_y + \frac{1}{2}\right) + \hbar\omega_z\left(n_z + \frac{1}{2}\right) \pm \frac{1}{2} g' \mu_B B, \tag{S4}$$

i.e., Eq. (1) in the main text. The classification is visually represented using different colors and line types. Notably, a key principle guiding this classification is that subbands belonging to the same group will never cross each other. Additionally, two subbands will intersect no more than once, consistent with the harmonic oscillator model. Given that the observed band crossings involve at most 5 subbands at high magnetic field and gate voltage (chemical potential), corresponding to the quantized plateau index increasing from 10 to 15, we categorize the subbands into 5 different classes. Each class is characterized by a unique quantum number $n_z$, resulting from the size quantization in the $z$ direction.

To interpret the distinctive subband evolution observed in our experiments, we begin with Eq. (1) in the main text, which can be reduced to a linear polynomial in terms of $n_y$ and $n_z$ as:

$$E(n_y, n_z) = E_0 + \eta n_y + \zeta n_z, \tag{S5}$$

where $\eta = \hbar\Omega = \hbar\sqrt{\omega_y^2 + (eB/m')^2}$, $\zeta = \hbar\omega_z$, and $E_0 = (\hbar\Omega + \hbar\omega_z)/2$.

In Eq. (S5), the Zeeman splitting term has been omitted, as its effect is barely observable in our measurements due to the strong suppression of the renormalized effective $g$-factor arising from the hidden Rashba effect and interlayer hybridization. Notably, the coefficient $\eta$ can be tuned via the magnetic-field-induced modulation of the confining potential in $y$ direction, while the coefficient $\zeta$ remains unchanged. This tunability leads to different subband degeneracy when varying magnetic fields, which is manifested as subband crossings in Fig. S10. The degeneracy at a certain energy $\mu_0$ is actually the number of solutions of Eq. (S5), depending on the values of $\eta$, $\zeta$, and $\mu_0$.

In Table SII, we summarize four representative types of subband crossings corresponding to $\eta/\zeta = 1$, $3/4$, $2/3$, and $1/2$. In particular, when $\eta/\zeta = 1$ around 10 T (Fig. S10), the degeneration degree ($q$) of the subbands follows an integer sequence $1, 2, 3, 4, \ldots$, whose cumulative sum yields a sequence that corresponds to Pascal triangle series $1, 3, 6, 10, \ldots$ with the transverse dimension of 2. Note that for the data shaded in blue, the corresponding subbands (dashed lines) are not plotted in Fig. S10, S11 due to the crowd. Based on Table SII, the subband crossings with a high degeneration degree can be labeled, as shown in Fig. S12.

## I. Extraction of the energy resolution

In this section, extraction of the energy resolution is presented. Based on the d.c.-bias spectroscopy (Fig. S13), by fitting the transconductance trace under magnetic field $B$ = 8 T using a Gaussian distribution function, the half width at half maximum (HWHM) yields a resolution of $E_{\mathrm{res}} = 2\sqrt{\ln 2} \cdot \Gamma = 0.83$ meV, corresponding to a broadening of $\Gamma = 0.7$ meV. As a result, this energy resolution $E_{\mathrm{res}}$ provides an upper bound on the effective $g$-factor ($g_{\mathrm{sup}}$). At the maximum $B$ = 12 T where the subband splitting due to the Zeeman effect remains undetectable, $g_{\mathrm{sup}} = 1.18$, following:

$$E_{\mathrm{res}} = g_{\mathrm{sup}} \cdot \mu_B \cdot B, \tag{S6}$$

where $\mu_B$ is the Bohr magneton. Notably, since such analysis gives an upper bound, the effective $g$-factor should be smaller than $g_{\text{sup}}$. Next, we extract the renormalized effective $g$-factor from the ab initio model involving the hidden Rashba effect.

## J. Effective model and simulation

This section presents the construction of the hidden Rashba bilayer model and the methodology for extracting the renormalized effective mass $m'$ and effective $g$-factor $g'$. By considering interlayer coupling of electrons between the Rashba monolayer with alternating Rashba spin-orbit coupling (SOC) strength $+\alpha$ and $-\alpha$, we derive the tight-binding Hamiltonian for the hidden Rashba model [47]:

$$H_{ll'} = \frac{p_x^2 + p_y^2}{2m^*}\delta_{ll'}\sigma_0 - t\delta_{l,l'\pm 1}\sigma_0 + (-)^l \frac{\alpha}{\hbar}\left(p_y\sigma_x - p_x\sigma_y\right)\delta_{ll'} + \frac{1}{2}g_e\mu_B B\delta_{ll'}\sigma_z, \tag{S7}$$

where $p_i = -i\hbar\partial_i + eA_i$ $(i = x, y)$ are canonical momenta, $t$ denotes the interlayer hybridization, and $l$ labels the layer index. The interlayer coupling strength $t = \hbar^2/(2m^*c^2)$ can be determined by the renormalized effective mass $m^*$ and the lattice constant along the $c$-axis $c$ = 1.216 nm. To explicitly incorporate Landau quantization, we apply an exact unitary transformation:

$$\begin{cases} a = \dfrac{\ell_B}{\sqrt{2}\hbar}\left(p_x + ip_y\right) \\ a^\dagger = \dfrac{\ell_B}{\sqrt{2}\hbar}\left(p_x - ip_y\right) \end{cases}, \tag{S8}$$

where $a$ and $a^\dagger$ are the annihilation and creation operators of Landau levels, satisfying the commutative relation $[a, a^\dagger] = 1$, and $\ell_B = \sqrt{\hbar/eB}$ is the magnetic length. Thus, under the Landau level basis, the Hamiltonian matrix elements become:

$$\begin{aligned} H_{nn'}^{ll'} = \hbar\frac{eB}{m^*}\left(n + \frac{1}{2}\right)\delta_{nn'}\delta_{ll'}\sigma_0 + \frac{1}{2}g_e\mu_B B\delta_{nn'}\delta_{ll'}\sigma_z - t\delta_{nn'}\delta_{l,l'\pm 1}\sigma_0 \\ +(-1)^l\frac{-i\alpha}{\sqrt{2}\ell_B}\,\delta_{ll'}\left(\sqrt{n+1}\,\delta_{n,n'-1}\sigma_+ - \sqrt{n}\,\delta_{n,n'+1}\sigma_-\right). \end{aligned} \tag{S9}$$

By taking $\alpha = 1.45$ eV · Å [28] and considering a 40-unit-cell-thick $Bi_2O_2Se$ nanoribbon (i.e. 80 monolayers, corresponding to the thickness of 50 nm), we find that $m^* = 0.08\, m_e$ and correspondingly $t = 0.32$ eV provide a good match to the experimental data shown in Fig. 4b. We then numerically diagonalize the Hamiltonian, including the lowest 40 Landau levels. This cut off ensures computational efficiency while maintaining sufficient accuracy, as higher Landau levels are well separated in energy and have negligible contribution to the low-energy behavior. From the calculated Landau levels of the hidden Rashba bilayer model, as illustrated in Fig. S14, we can extract the renormalized g-factor $g' = 0.78$ based on the fact that the effective Zeeman splitting $E_z = g'\mu_B B$ is proportional to the magnetic field. This strongly suppressed renormalized effective $g$-factor emerges from the hidden Rashba effect, where the interlayer coupling mixes the opposite spin textures in the monolayers with alternating Rashba SOC strength, further resulting in the suppression of the effective Zeeman splitting under a high magnetic field. In addition, the Landau levels exhibit linear dependence on $B$. From the slope of the linear dispersion, the renormalized effect mass $m' = 0.084m_e$ is determined according to $E_n = \hbar\frac{eB}{m'}\left(n+\frac{1}{2}\right)$.

Subsequently, to obtain the subband evolution under the confining potentials in both $y$ and $z$ directions, we demonstrate an effective model, where electrons, with the renormalized effective mass $m'$ and $g$-factor $g'$, are confined under a magnetic field $B$, taking the gauge of $\boldsymbol{A} = A_x\,\boldsymbol{e}_x = -By\,\boldsymbol{e}_x$ and the harmonic oscillator potentials with the natural frequency of $\omega_y$ and $\omega_z$ in the $y$ and $z$ directions, respectively. By solving the eigenenergies of the model, the subband bottom is given by Eq. (S4). Further, the conductance $G$ is proportional to the number of states nearby the Fermi levels, i.e.,

$$G(\mu,B) = \frac{e^2}{h}\sum_{n_y,n_z} H\left(\mu - E\left(n_y,n_z;B\right)\right), \tag{S10}$$

where $H(x)$ is the Heaviside function. Moreover, considering the energy broadening from the unavoidable effects of finite temperature and measurement noise, the observed conductance is blurred by an assumed Gaussian broadening $W(\mu;\varepsilon_0,\Gamma) = \frac{1}{\sqrt{2\pi}\Gamma}\,e^{-(\mu-\varepsilon_0)^2/(2\Gamma^2)}$ with $\Gamma = 0.7$ meV. Under such scenario,

$$\begin{aligned} G(\mu,B) &= \frac{e^2}{h}\int\sum_{n_y,n_z} W\left(\mu;E\left(n_y,n_z;B\right),\Gamma\right)\mathrm{d}\mu \\ &= \frac{e^2}{h}\cdot\sum_{n_y,n_z}\frac{1}{2}\mathrm{erf}\left(\frac{\mu - E\left(n_y,n_z;B\right)}{\sqrt{2}\Gamma}\right), \end{aligned} \tag{S11}$$

where $\mathrm{erf}(x) = \frac{2}{\sqrt{\pi}}\int_0^x \exp(-t^2)\,\mathrm{d}t$ is the error function. Therefore, the conductance and transconductance maps as a function of $B$ and $\mu$ can be obtained.

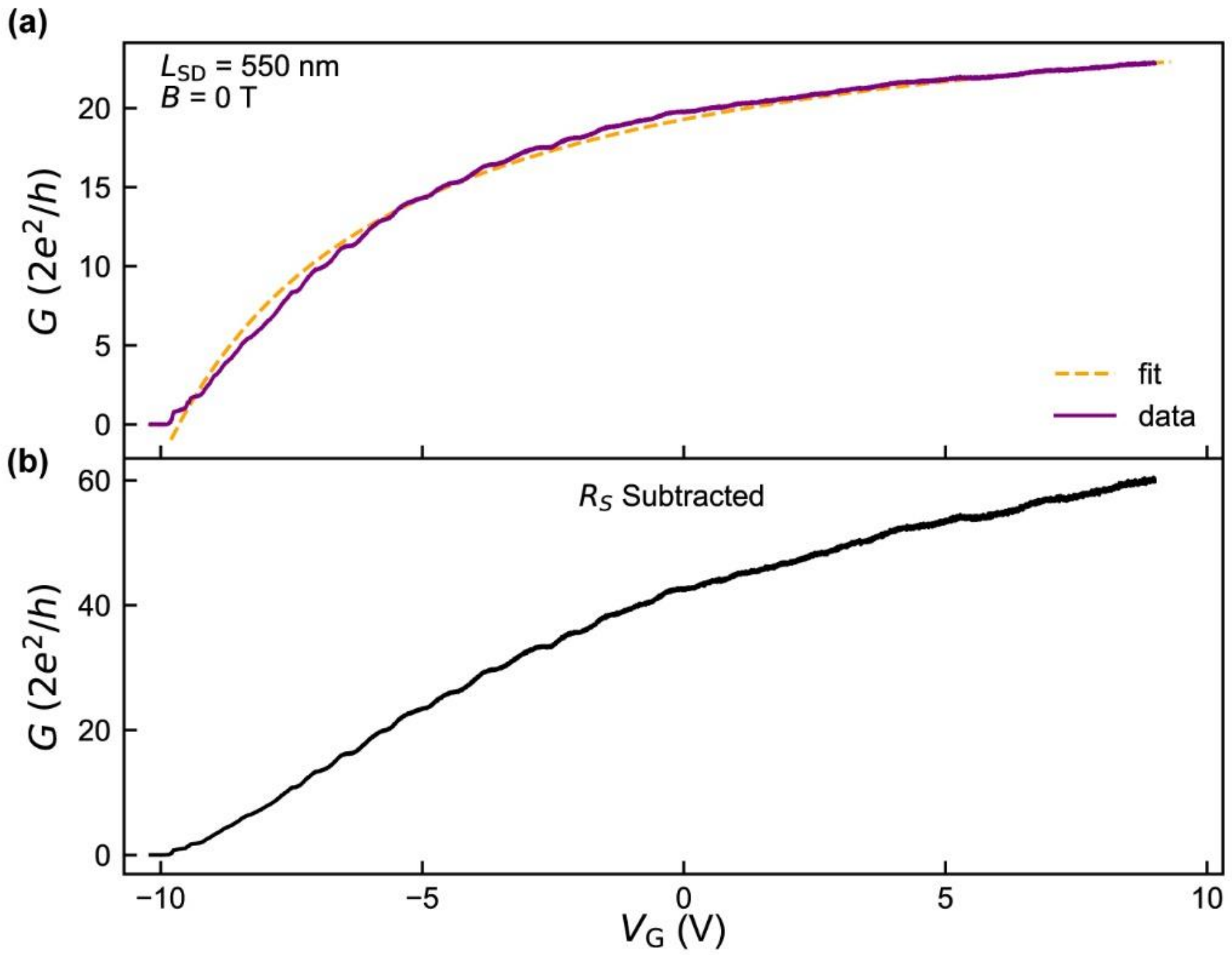

FIG. S1. Conductance measured at $B$ = 0T. (a) Measured conductance $G$ as a function of gate voltage $V_G$ under zero magnetic field (solid curve). The orange dashed line represents a fitting of the data using Eq. (S1). (b) The same data as in (a), but with a series resistance $R_S = 350\ \Omega$ subtracted.

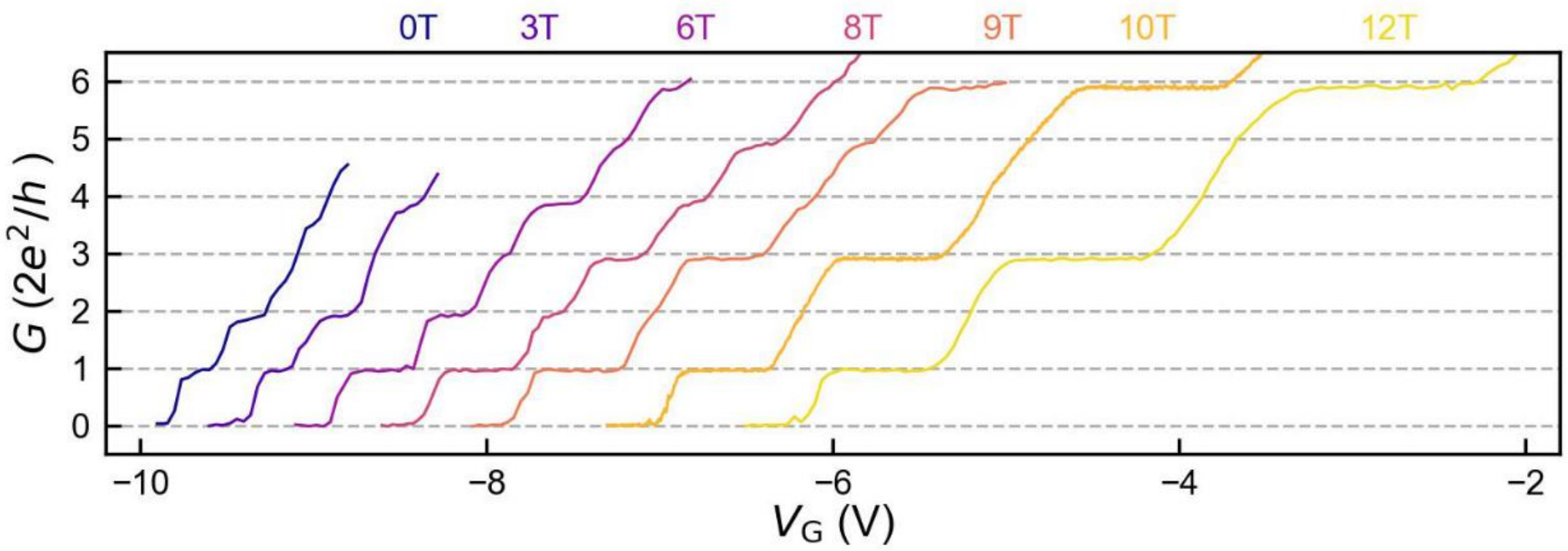

FIG. S2. Quantized conductance at different magnetic fields. Conductance traces, with the series resistance of $R_s$ = 350 Ω subtracted, as a function of gate voltage $V_G$, under different magnetic fields. For clarity, the curves are offset by 0.45 V (below 9 T) and 0.8 V (above 9 T).

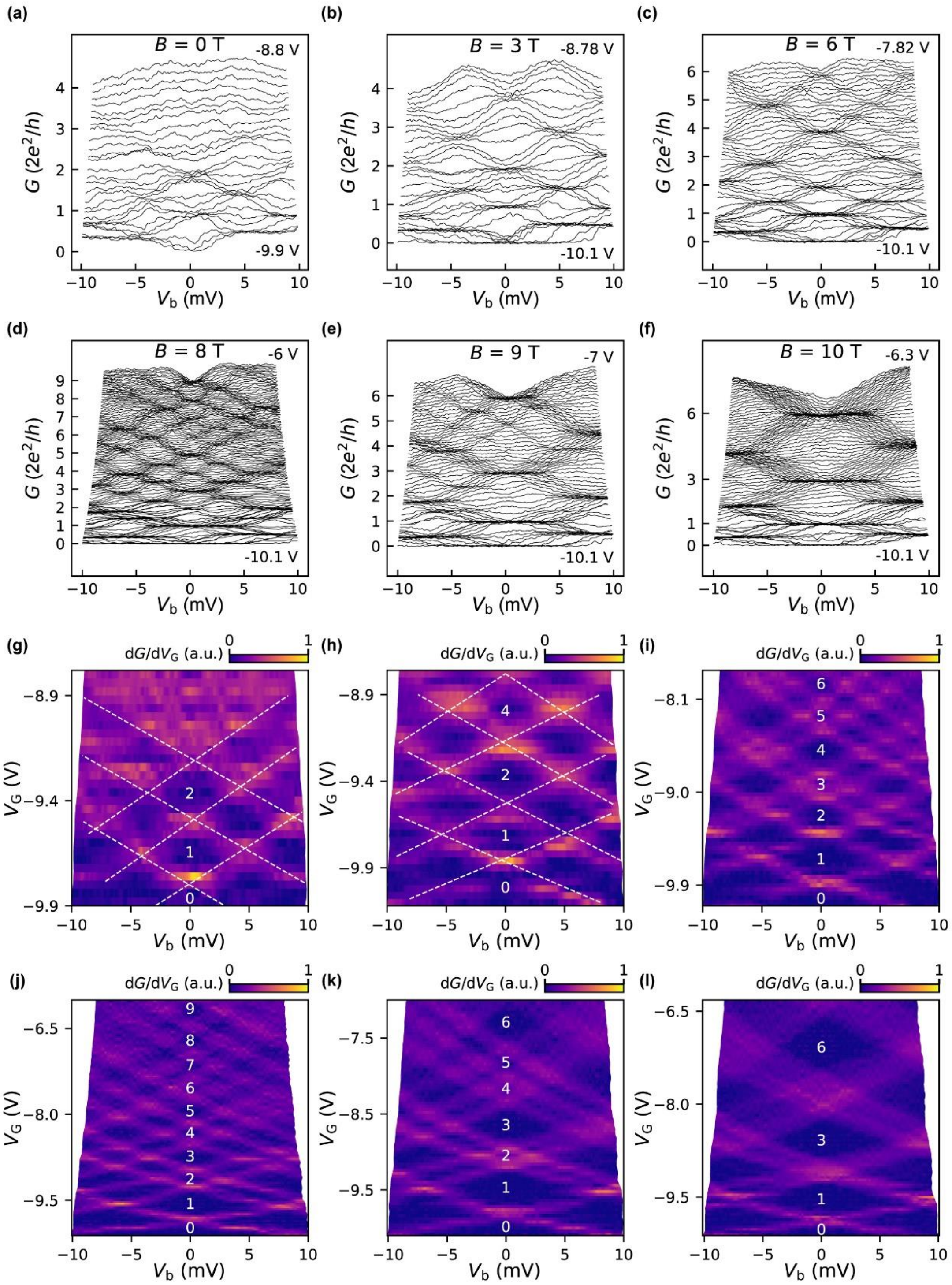

FIG. S3. Bias spectroscopy at different magnetic fields. (a-f) Conductance $G$ as a function of bias voltage $V_{\mathrm{b}}$. (f-l) Corresponding transconductance spectra of (a-f). The numbers label the quantized conductance plateaus. (a), (g) $B$ = 0 T. (b), (h) $B$ = 3 T. (c), (i) $B$ = 6 T. (d), (j) $B$ = 8 T. (e), (k) $B$ = 9 T. (f), (l) $B$ = 10 T.

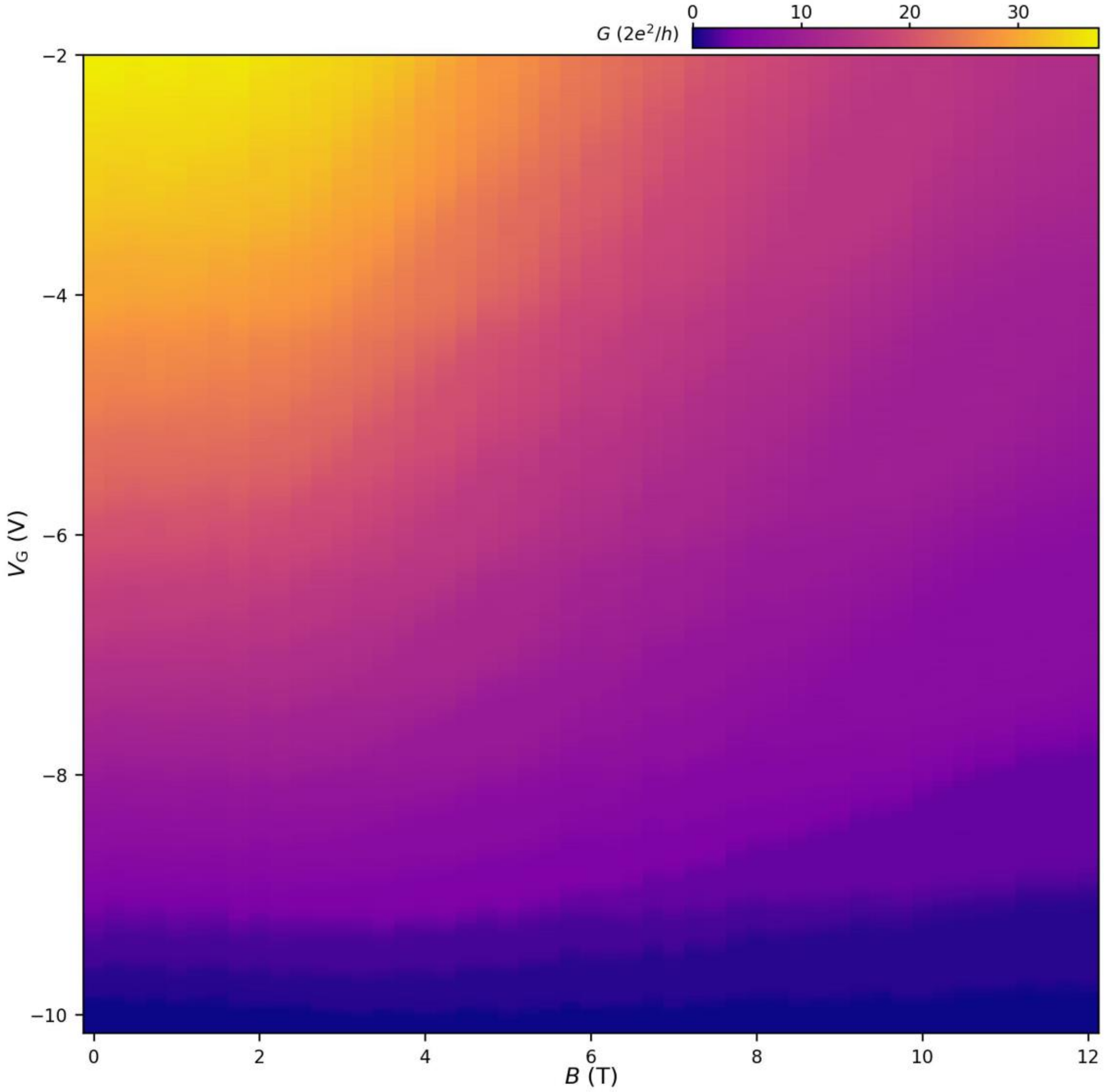

FIG. S4. Evolution of quantized conductance plateaus. Conductance map plotted against gate voltage ($V_{\mathrm{G}}$) and magnetic field ($B$).

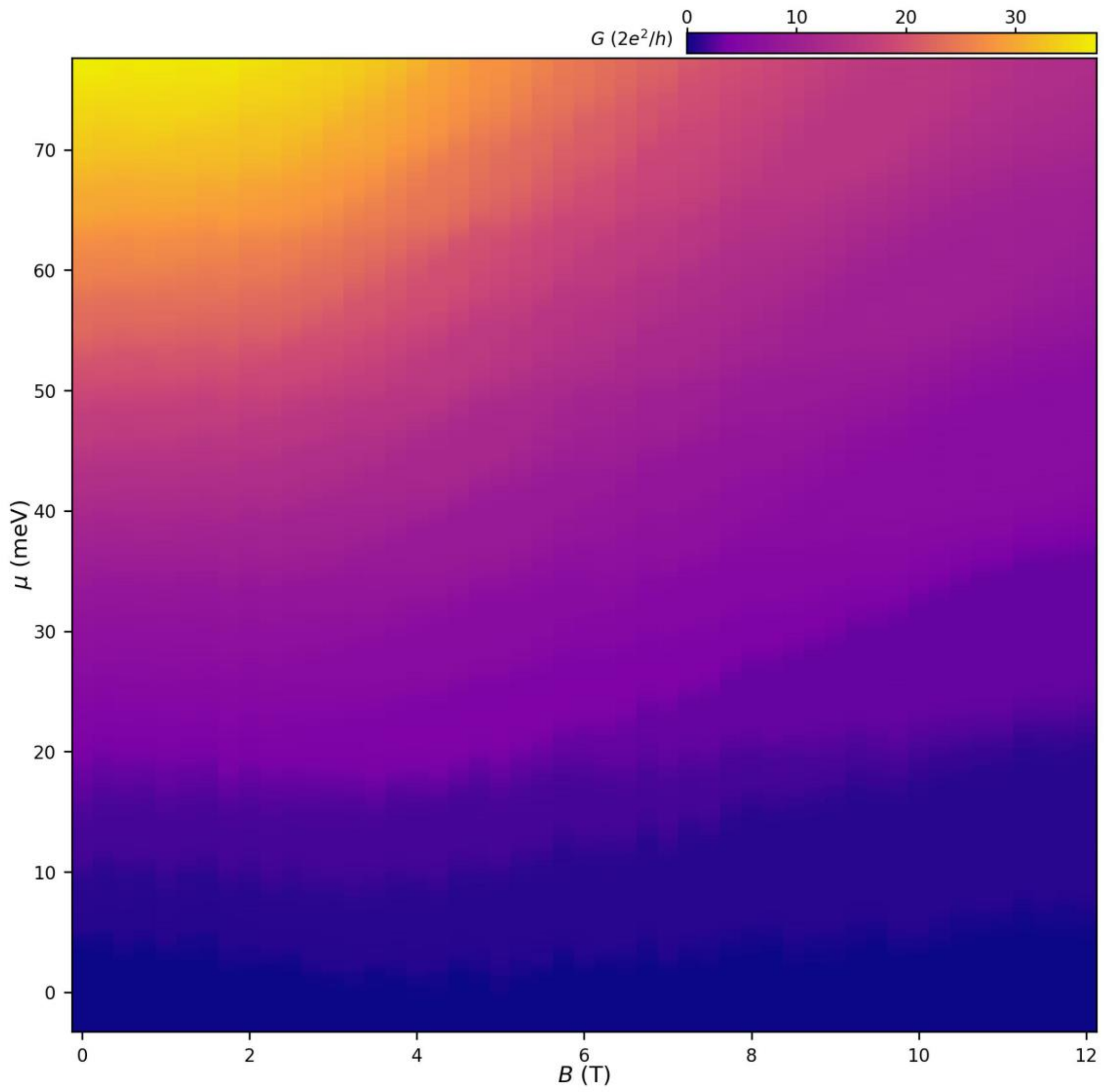

FIG. S5. Evolution of quantized conductance plateaus. Conductance map as a function of chemical potential ($\mu$) and magnetic field ($B$). The transformation from gate voltage to chemical potential will be shown later.

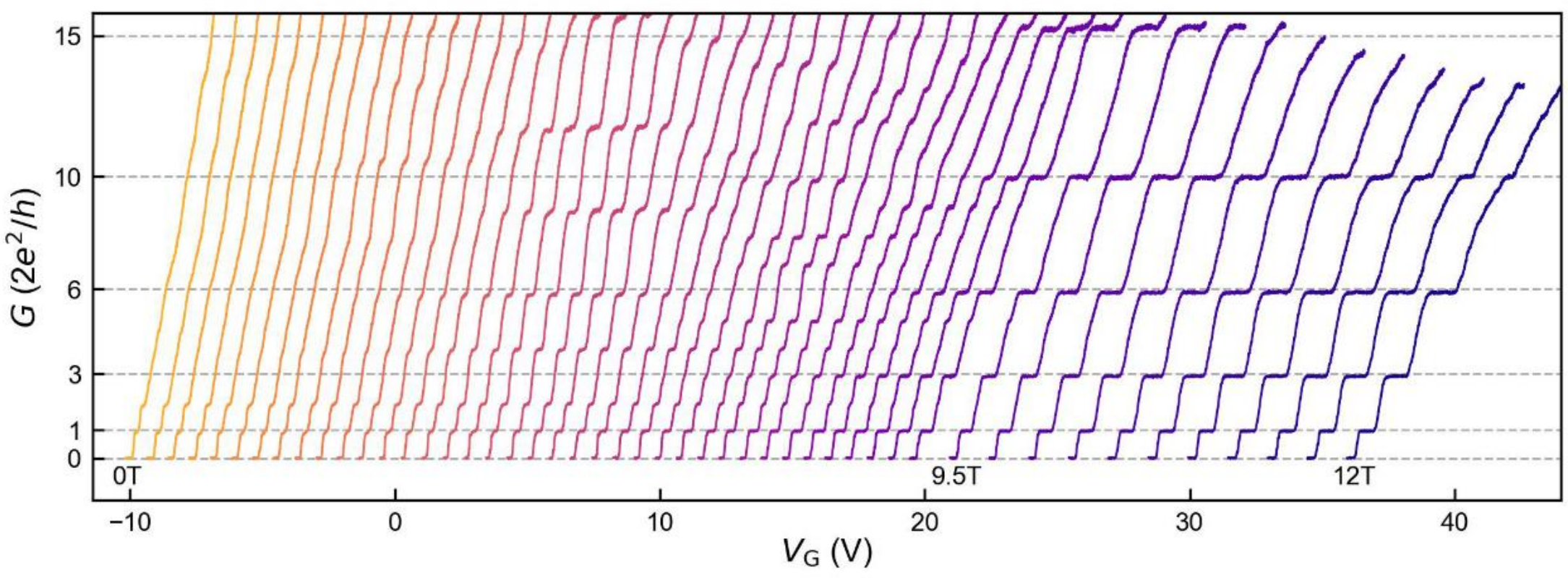

FIG. S6. Evolution of quantized conductance plateaus. Conductance traces plotted as a function of gate voltage ($V_{\mathrm{G}}$). For clarity, the curves are offset by 0.8 V (for B < 9.5 T) and 1.5 V (for B > 9.5 T). The emerge and disappearance of quantized plateaus can be recognized.

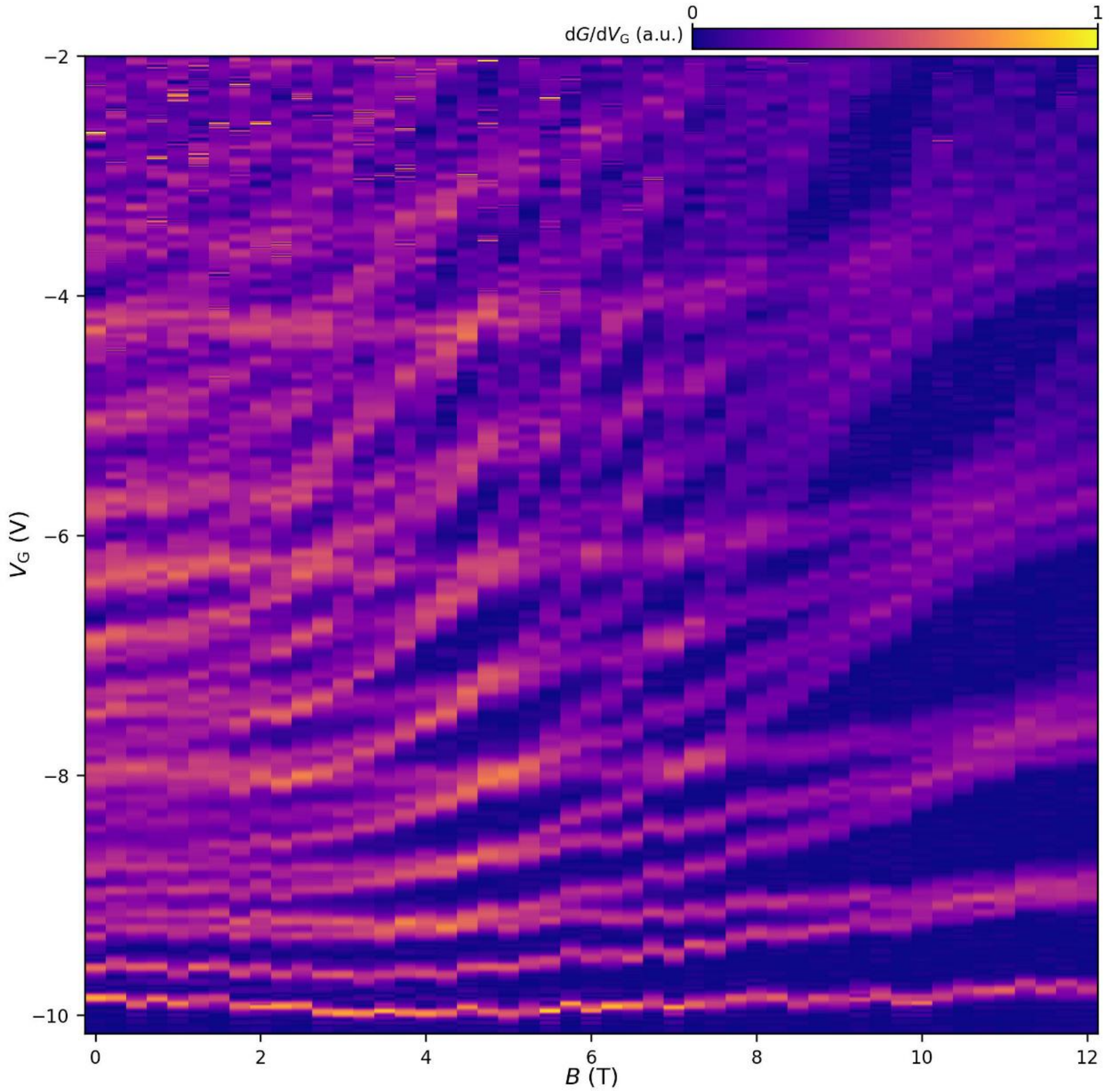

FIG. S7. Evolution of quantized conductance plateaus. Enlarged transconductance map illustrating the dependence on gate voltage ($V_G$) and magnetic field ($B$).

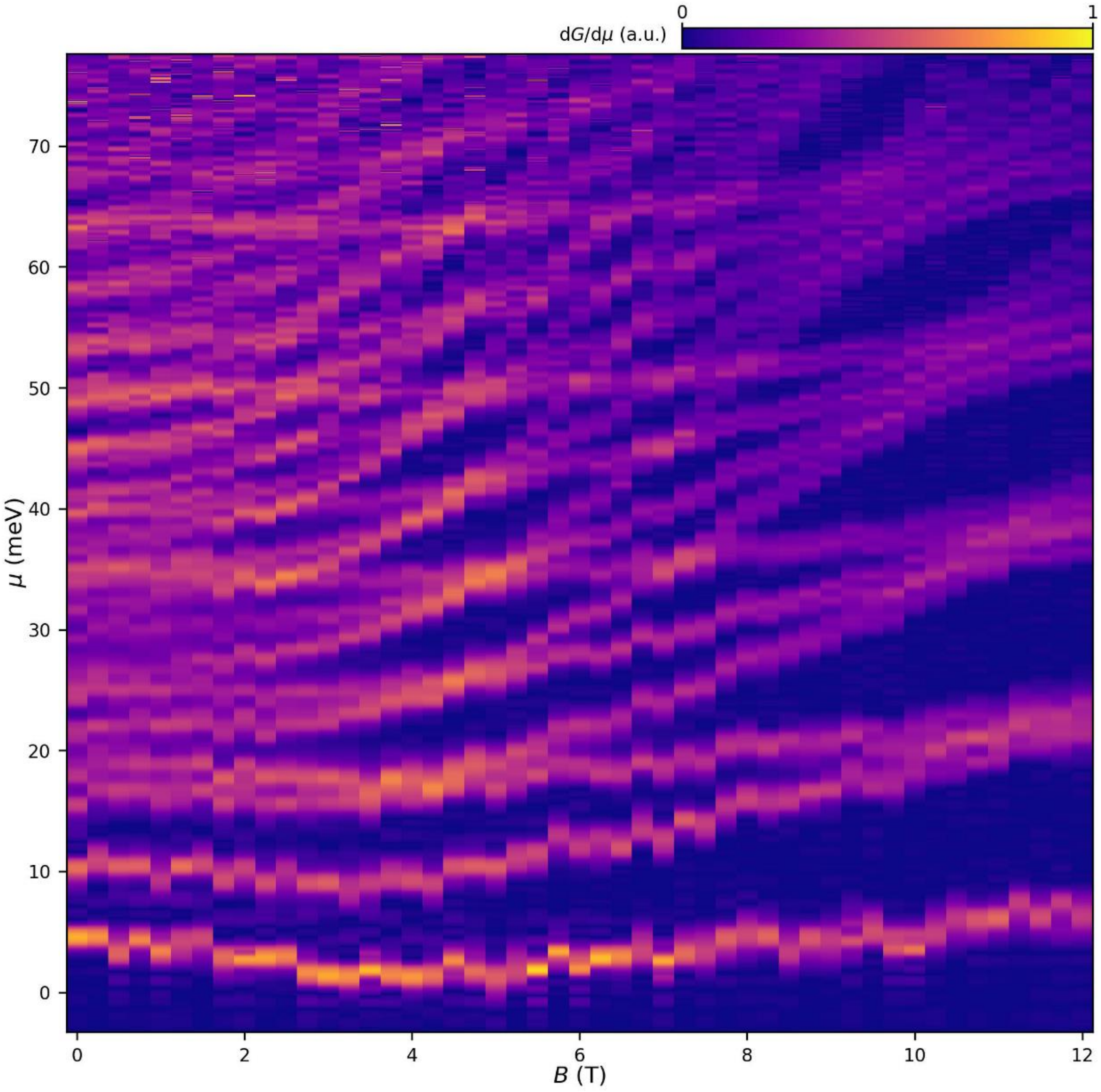

FIG. S8. Evolution of quantized conductance plateaus. Enlarged transconductance map as a function of chemical potential ($\mu$) and magnetic field ($B$), which is the same plot as shown in Fig. 4(b).

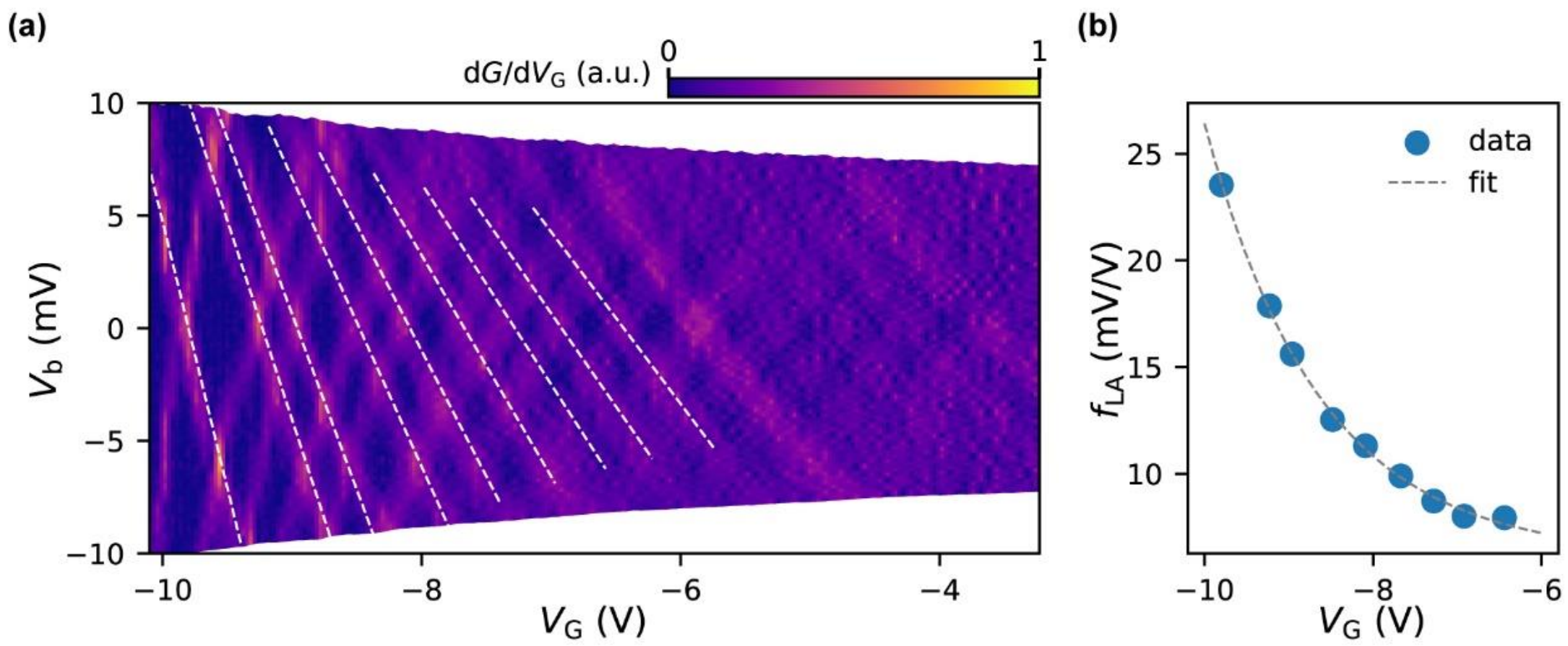

FIG. S9. Transformation from gate voltage to chemical potential. (a) Transconductance map as a function of d.c. bias and gate voltage at $B = 8$ T. Dashed lines indicate subband slopes used to quantify the gate lever arm. (b) Fitting of the gate lever arm with respect to $V_{\mathrm{G}}$.

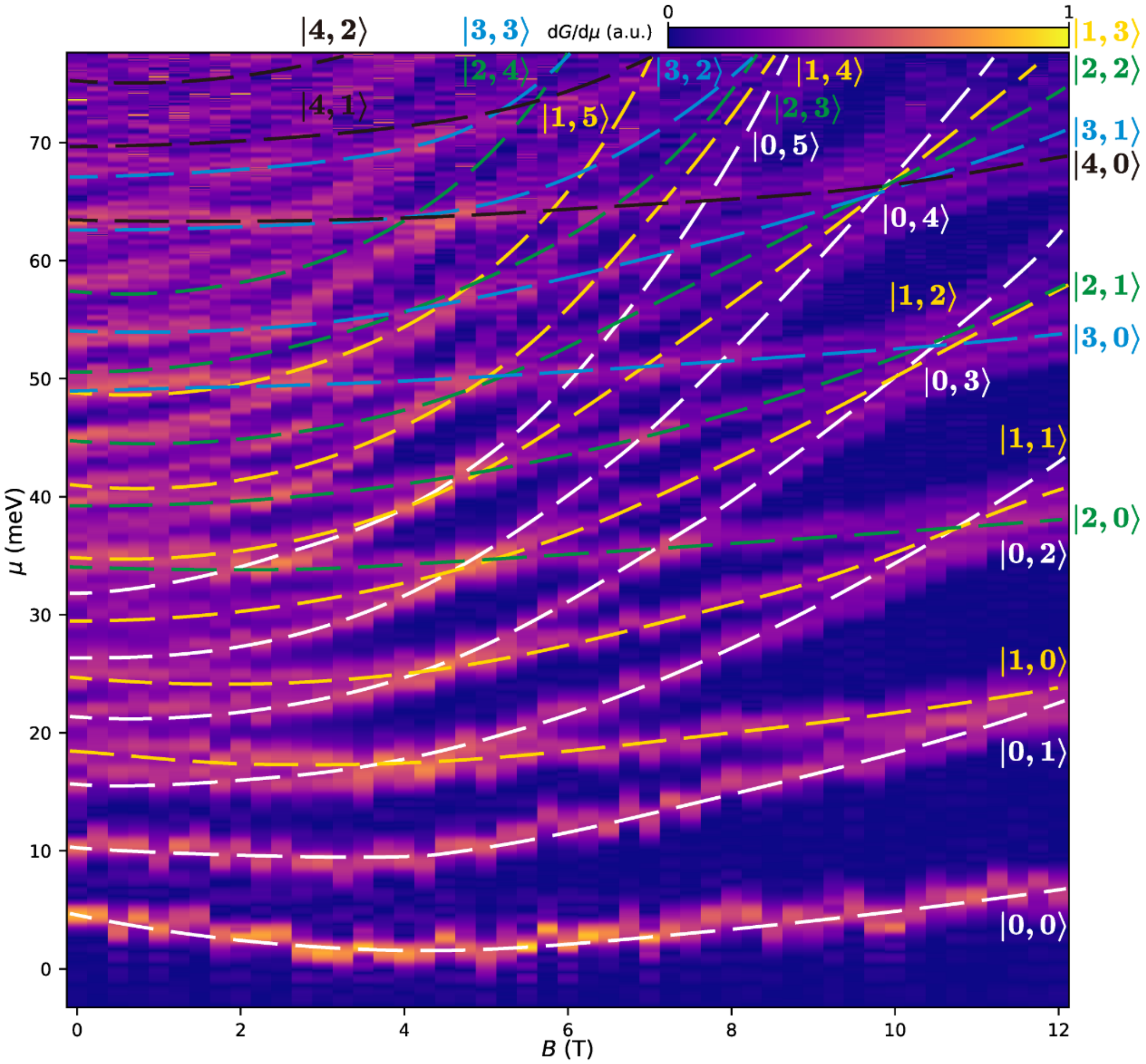

FIG. S10. Evolution of quantized conductance plateaus. Transconductance map illustrating the subband evolution, with individual subbands denoted by dashed lines and labels.

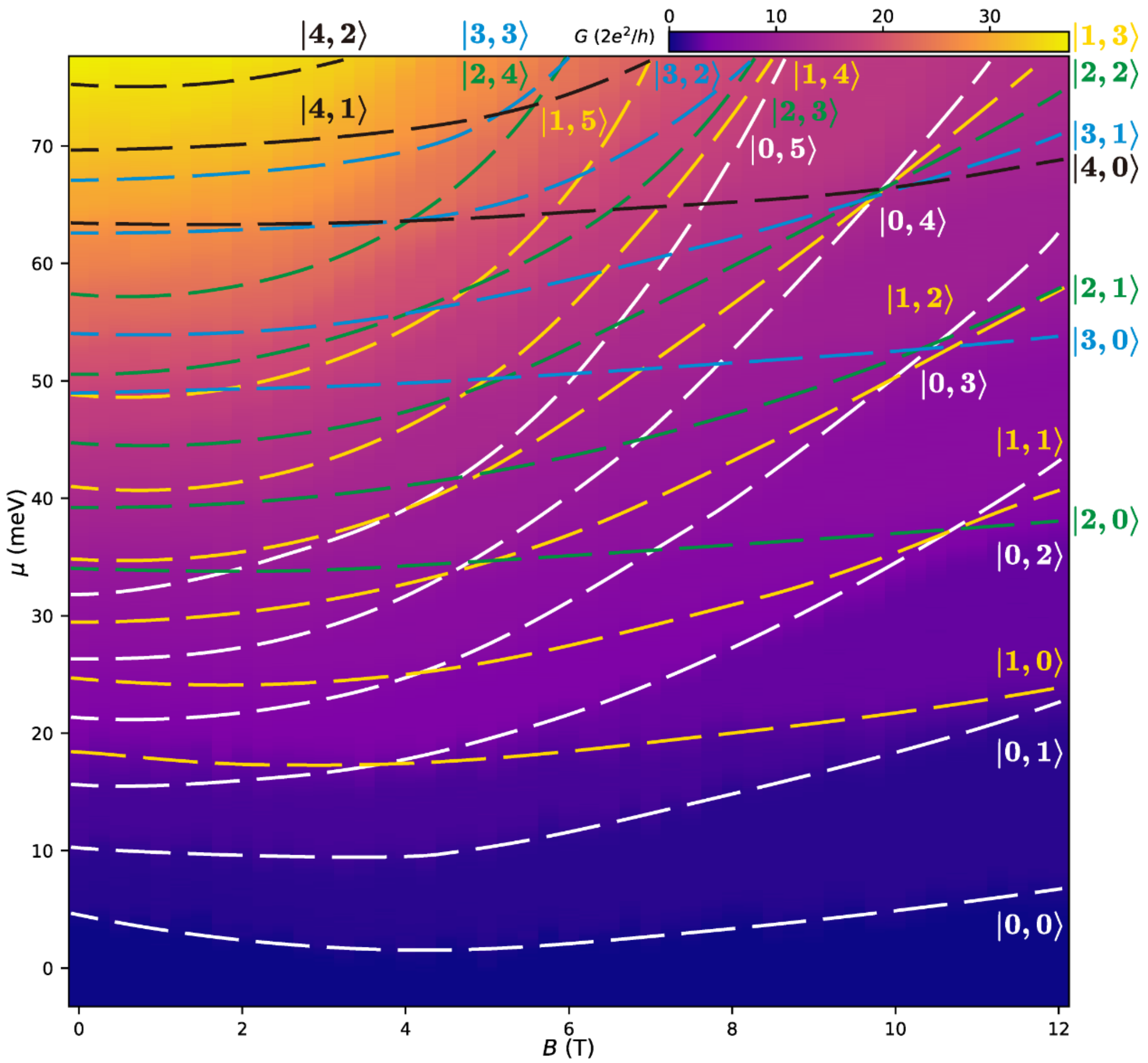

FIG. S11. Evolution of quantized conductance plateaus. Conductance map illustrating the subband evolution, with individual subbands denoted by dashed lines and labels which are the same as in Fig.S10.

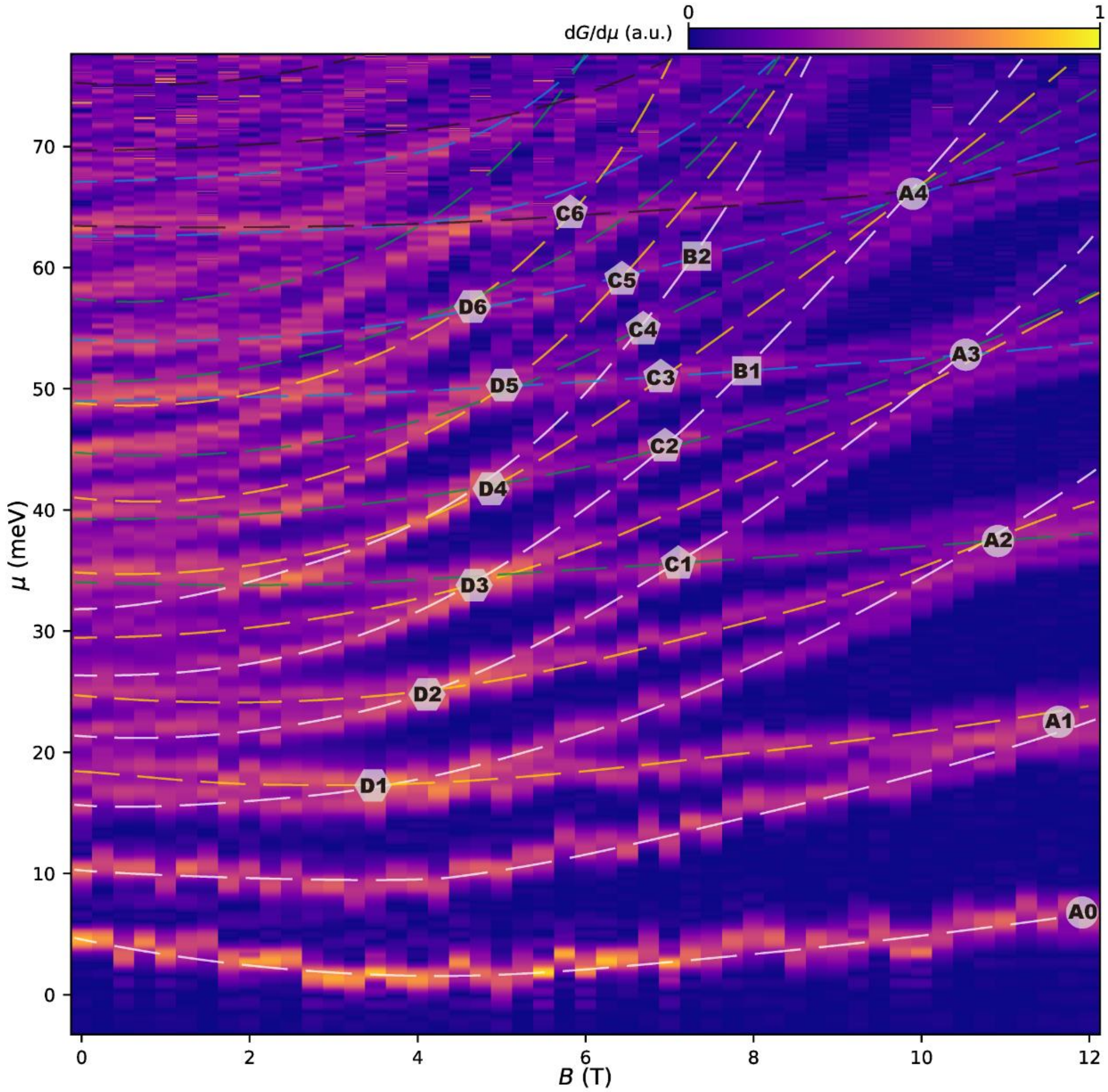

FIG. S12. Evolution of quantized conductance plateaus. The same as Fig. S10, where subband crossings are represented by different types of labels according to Table SII.

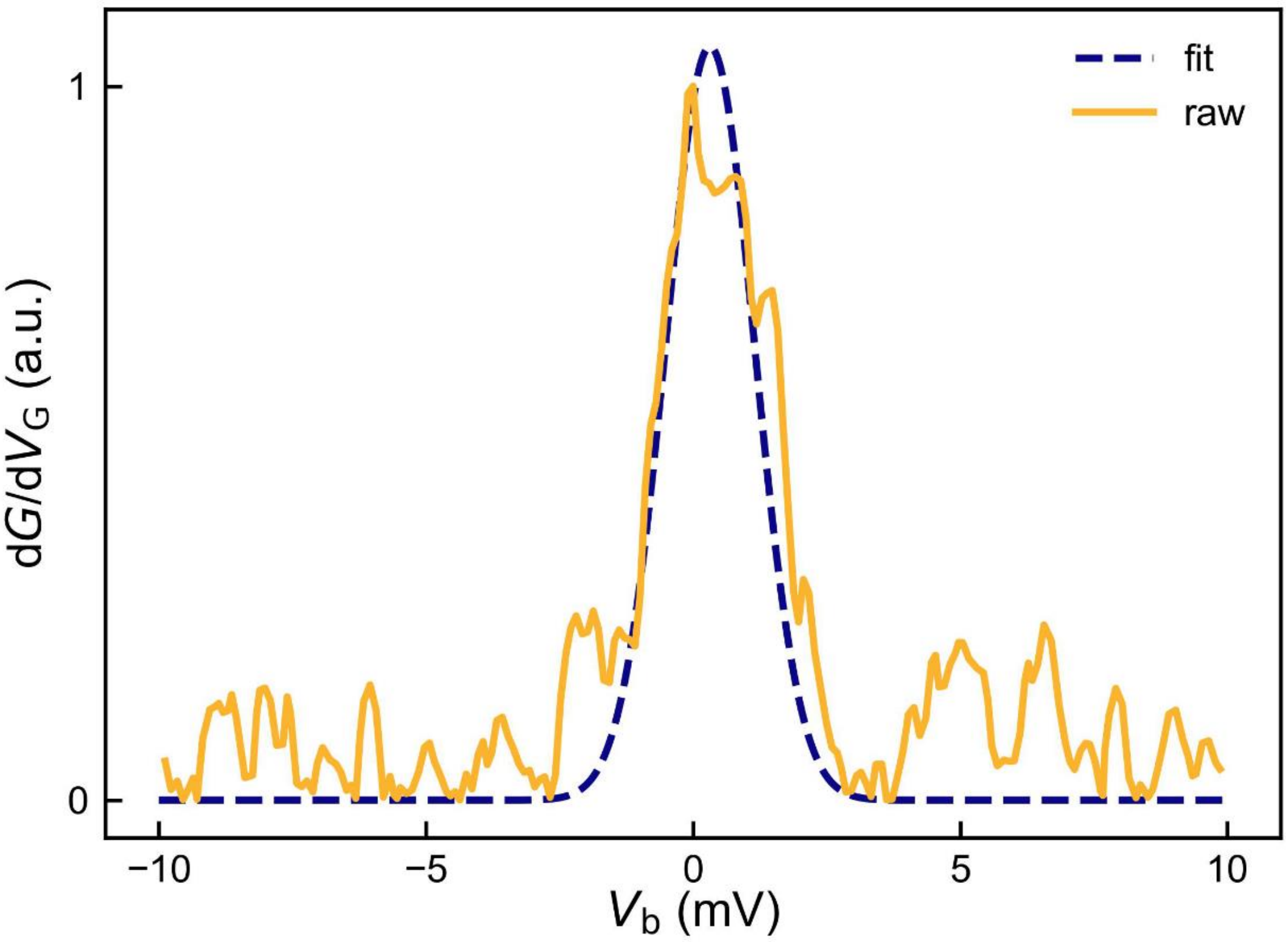

FIG. S13. Extraction of the energy resolution. Fitting the transconductance curve at $B$ = 8 T, $V_G$ = -9.86 V (orange solid line) with Gaussian distribution (purple dashed line) gives a resolution of 0.83 meV.

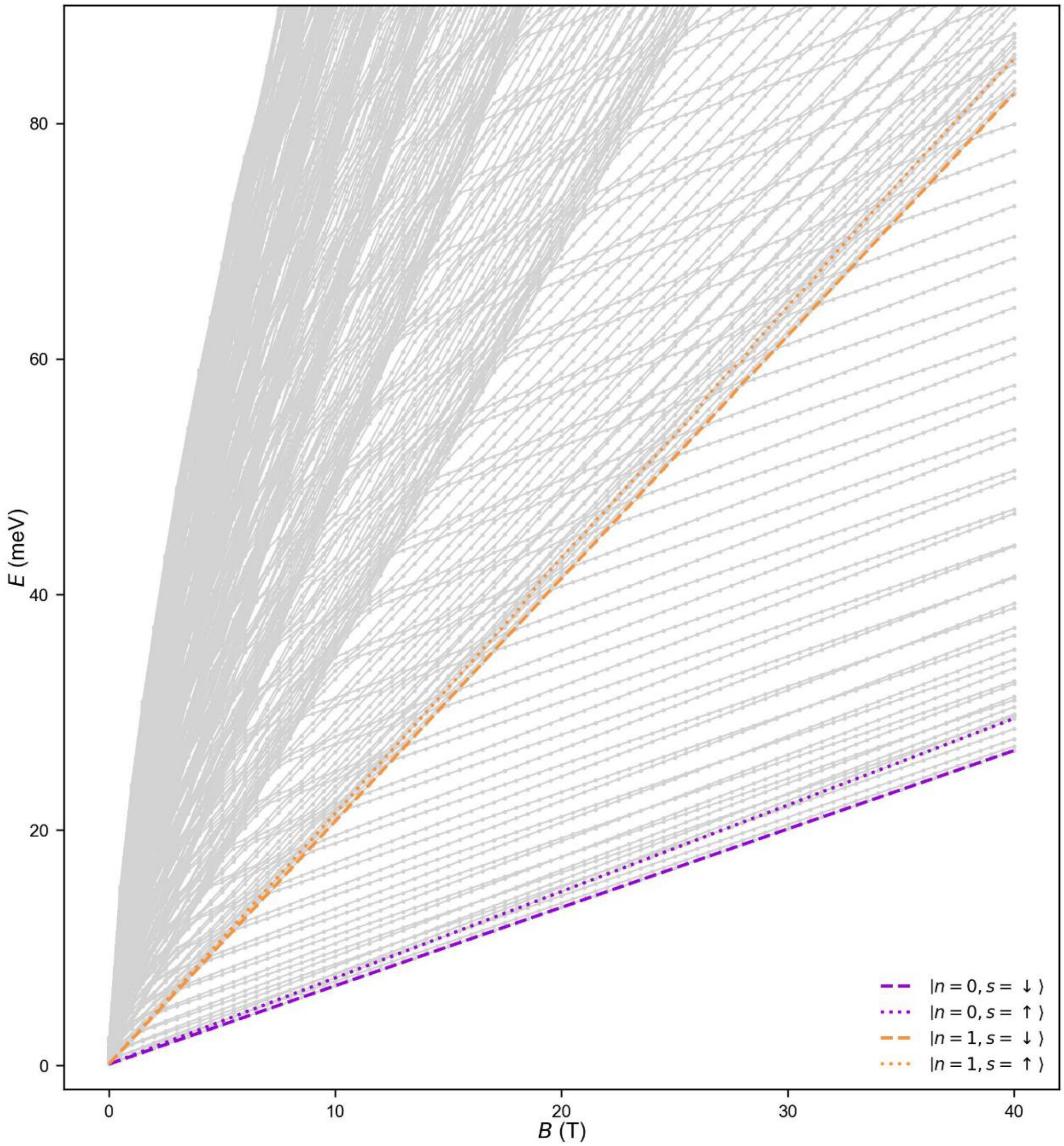

FIG. S14. Landau levels of the hidden Rashba bilayer model as a function of the magnetic field $B$. As the magnetic field increases, the Landau levels fan out, exhibiting energy splitting due to Zeeman effect and dimension quantization in $z$ direction. For clarity, only the lowest 300 levels are shown. The violet dashed and dotted lines denote the lowest Landau level with index $n = 0$ for spin-up and spin-down states, respectively, while the orange lines correspond to the $n = 1$ Landau level.

TABLE SI. Reported quantized plateaus at $B = 0$ T in individual 1D materials.

| Material class | Compound | Plateau index | Reference |
|---|---|---|---|
| Semiconductors | $Bi_2O_2Se$ | 44 | **This work** |
| | CNT | 3 | 49 |
| | PbTe | 2 | 50 |
| | | 3 | 51 |
| | | 4 | 52 |
| | | 3 | 53 |
| | InAs | 4 | 54 |
| | | 2 | 55 |
| | | 4 | 56 |
| | | 3 | 57 |
| | | 2 | 58 |
| | | 1 | 59 |
| | | 3 | 62 |
| | InSb | 2 | 16 |
| | | 5 | 60 |
| | | 5 | 62 |
| | | 1 | 63 |
| | | 1 | 64 |
| | | 2 | 65 |
| | Ge/Si | 5 | 17 |
| | Al-Ge-Al | 3 | 66 |
| Metals | Cu | 2 | 67 |
| | Pt | 5 | 68 |
| | O/Cu | 6 | |
| | Cu | 5 | |
| | Au | 6 | 69 |
| | | 10 | 70 |

TABLE SII. Summarization of the subband crossings.

| | $q$ | $n_y$ | $n_z$ |
|---|---|---|---|
| $\eta/\zeta = 1$ | | | |
| **A0** | 1 | 0 | 0 |
| **A1** | 2 | 0 | 1 |
| | | 1 | 0 |
| **A2** | 3 | 0 | 2 |
| | | 1 | 1 |
| | | 2 | 0 |
| **A3** | 4 | 0 | 3 |
| | | 1 | 2 |
| | | 2 | 1 |
| | | 3 | 0 |
| **A4** | 5 | 0 | 4 |
| | | 1 | 3 |
| | | 2 | 2 |
| | | 3 | 1 |
| | | 4 | 0 |

| | $q$ | $n_y$ | $n_z$ |
|---|---|---|---|
| $\eta/\zeta = 3/4$ | | | |
| **B1** | 2 | 0 | 3 |
| | | 4 | 0 |
| **B2** | 2 | 1 | 3 |
| | | 5 | 0 |
| $\eta/\zeta = 2/3$ | | | |
| **C1** | 2 | 0 | 2 |
| | | 3 | 0 |
| **C2** | 2 | 1 | 2 |
| | | 4 | 0 |
| **C3** | 2 | 0 | 3 |
| | | 3 | 1 |
| **C4** | 2 | 2 | 2 |
| | | 5 | 0 |
| **C5** | 2 | 1 | 3 |
| | | 4 | 1 |
| **C6** | 3 | 0 | 4 |
| | | 3 | 2 |
| | | 6 | 0 |

| | $q$ | $n_y$ | $n_z$ |
|---|---|---|---|
| $\eta/\zeta = 1/2$ | | | |
| **D1** | 2 | 0 | 1 |
| | | 2 | 0 |
| **D2** | 2 | 1 | 1 |
| | | 3 | 0 |
| **D3** | 3 | 0 | 2 |
| | | 2 | 1 |
| | | 4 | 0 |
| **D4** | 3 | 1 | 2 |
| | | 3 | 1 |
| | | 5 | 0 |
| **D5** | 4 | 0 | 3 |
| | | 2 | 2 |
| | | 4 | 1 |
| | | 6 | 0 |
| **D6** | 4 | 1 | 3 |
| | | 3 | 2 |
| | | 5 | 1 |
| | | 7 | 0 |